\documentclass[floatfix,twocolumn, prb, longbibliography, superscriptaddress]{revtex4-2}
\pdfoutput=1
\usepackage{graphicx}
\usepackage{bbold}

\usepackage{subfigure}
\usepackage{float}
\usepackage{amsmath}
\usepackage{textcomp,gensymb} 

\usepackage[colorlinks=true,linkcolor=blue,citecolor=blue,filecolor=blue,urlcolor=blue]{hyperref}

\usepackage[english]{babel}

\begin{document}
\title{Influence of the electron spill-out and nonlocality on gap-plasmons in the limit of vanishing gaps}

\author{M. Khalid}
\email{muhammad.khalid@iit.it}
\affiliation{Istituto Italiano di Tecnologia, Center for Biomolecular Nanotechnologies, Via Barsanti 14, 73010 Arnesano, Italy.}
\author{O. Morandi}
\affiliation{Dipartimento di Matematica e Informatica U. Dini, University of Florence, via S. Marta 3, 50149, Florence, Italy}
\author{E. Mallet }
\affiliation{Clermont Universit\'e, Institut Pascal (IP), BP 10448, F-63000 Clermont-Ferrand, France}
\affiliation{CNRS, UMR 6602, IP, F-63171 Aubi\`ere, France}
\author{P. A. Hervieux }
\affiliation{Université de Strasbourg, CNRS, Institut de Physique et Chimie des Matériaux de Strasbourg, UMR 7504, F-67000 Strasbourg, France}
\author{G. Manfredi}
\affiliation{Université de Strasbourg, CNRS, Institut de Physique et Chimie des Matériaux de Strasbourg, UMR 7504, F-67000 Strasbourg, France}
\author{A. Moreau}
\affiliation{Clermont Universit\'e, Institut Pascal (IP), BP 10448, F-63000 Clermont-Ferrand, France}
\affiliation{CNRS, UMR 6602, IP, F-63171 Aubi\`ere, France}
\author{C. Cirac\`i}
\email{cristian.ciraci@iit.it}
\affiliation{Istituto Italiano di Tecnologia, Center for Biomolecular Nanotechnologies, Via Barsanti 14, 73010 Arnesano, Italy.}

\date{\today}

\begin{abstract}
We study the effect of electron spill-out and of nonlocality on the propagation of light inside a gap between two semi-infinite metallic regions.
We compare the predictions of a local response model taking into account only the spill-out, to the predictions of a quantum hydrodynamic model able to take both phenomena into account.
We show that only the latter is able to correctly retrieve the correct limit when the gap closes, while the local model suffers from undesirable features (divergence of the fields, overestimation of the losses).
Finally, we show that, to a certain extent, the correct results can be retrieved using a simple local approach without spill-out or conventional Thomas-Fermi approximation, but considering an effective gap width.
\end{abstract}

\maketitle

\thispagestyle{plain} 

\section{Introduction}\label{sec:introduction}
The possibility to design extremely miniaturized devices, even optical circuits, has fostered research in plasmonics for decades. The most recent advances in fabrication processes have led to metallic nanostructures where extremely narrow gaps play a central role \cite{baumberg2019extreme}.
Metal-insulator-metal (MIM) structures support a mode with a particularly large effective index called the {\em gap-plasmon} \cite{Wang2007Phys.Rev.B., Jung2009Phys.Rev.B.}. The insertion of MIM sections in finite systems can thus lead to resonances in exceptionally small volumes \cite{akselrod2014probing,hoang2015ultrafast}, while the quality factor of the resonance \cite{yang2012ultrasmall} and its absorption cross-section \cite{moreau2012controlled} are essentially preserved.
The insulator gaps can be so small (typically smaller than 1 nm \cite{sondergaard2012plasmonic}) that many questions arise concerning the accuracy of the electromagnetic description at such scales. Indeed, at these scales the classical local response approximation (LRA), based on the Drude model, reaches its limits \cite{Ciraci2012Science}, as the spatial dispersion due to the repulsion between electrons inside the metal has a direct impact on the optical response of the system. Electron-electron interactions can be taken into account in a simple way in the framework of the hydrodynamic theory (HT) with hard-wall boundary conditions (i.e., no electron spill-out is allowed), by adding the Thomas-Fermi (TF) electron pressure term \cite{Raza2011PhysRevB,Ciraci2013ChemPhysChem}.

When dealing with the near-atomic subnanometer length scales, it becomes inevitable to consider the quantum nature of electrons, and the quantum mechanical effects, such as electron spill-out and quantum tunneling, become important. The TFHT cannot address these phenomena properly and, therefore, more sophisticated theories are required for an accurate description of electron dynamics \cite{Teperik2013Phys.Rev.Lett., Stella2013Phys.Chem., Zhu2016Nat.Com.}. Ab-initio quantum mechanical approaches, such as time-dependent density functional theory (TD-DFT), provide an appealing solution to such problems although they are often computationally demanding, especially for large plasmonic structures \cite{Zuloaga2009, Ullrich2011, Aguirregabiria2018}.

Within the LRA framework it is still possible to account for the tunneling by interposing between the metal regions a fictitious material described by an effective conductivity \cite{Esteban2012NatCommun} or  by using a finite-current boundary condition at the metal-insulator interface \cite{Feiginov1998JETPLett, Ryzhii2001JJAppPhys, Svintsov2016PhysRevB}.
However, when the LRA is used to account for spill-out, non-physical features (including a divergence of the associated electric field) arise \cite{SkjolstrupPhys.Rev.B.,Skjolstrup2019_PRB}. Due to this artificial divergence of the field, the imaginary part of the propagation constant of the gap-plasmon can be largely overestimated, which consequently results in very high mode-propagation losses.

 In the past few years, a hybrid approach called quantum hydrodynamic theory (QHT)\textemdash which adds a $\nabla n$-dependent correction ($n$ being the charge density) to the TF kinetic energy\textemdash has been developed to study nonlocal and quantum effects (electron spill-out/tunneling) in plasmonic systems more accurately \cite{Toscano2015NatCommun, Yan2015Phys.Rev.B, Ciraci2016Phys.Rev.B, Ciraci2017Phys.Rev.B, Ding2017PRB}. The QHT method intrinsically account for the electron spill-out and tunneling, but also spatial dispersion (nonlocality) which is relevant in the unidirectional plasmonic waveguides \cite{Gangaraj2019Optica} and nonlinear optical dynamics \cite{Gangaraj2020PhysRevApp, Khalid2020commphys}. This approach is based on orbital-free description of quantum electronic systems and its accuracy relies on the exactness of non-interacting kinetic energy functional. Ideally, if the exact expression of such functional were known, QHT would be as accurate as TD-DFT \cite{Runge.1984}.
 
Here we apply the QHT approach to the gap-plasmon in the vanishing gap limit, and compare to the LRA model in the presence of spill-out. We show that the unphysical divergences predicted by the LRA lead to a large and artificial overestimation of the absorption when the gap-plasmon propagates, but that including spatial dispersion in the description of the electron gas makes such features disappear. In fact, the gap-mode propagation constant is influenced by the presence of longitudinal modes (due to nonlocality) inside the metallic region, which are determined by the hydrodynamic equations and, in particular, by the quantum pressure term. The correct limit, as the gap closes, is thus retrieved only when both the spill-out and nonlocality are taken into account.

 In section \ref{sec:theory}, we detail the theoretical background of our study, first by deriving simplified equations corresponding to the LRA approximation in the presence of spill-out and then by detailing the differences with the QHT framework. In section \ref{sec:results}, we discuss our numerical results by comparing the behavior of the effective index for the different approaches of interest (LRA, TFHT, LRA with spill-out and QHT). Finally, in section \ref{sec:effective},
we discuss the implications of our work regarding the definition of the actual optical edges of a metal. We show that it makes sense to define an effective gap for the gap-plasmon, allowing to understand why in so many works LRA without spill-out has been so successfully applied.

\section{Theoretical framework}\label{sec:theory}
The propagation of the electric and magnetic fields, $\bf E$ and $\bf H$, of the gap-plasmon is described by Maxwell's equations:
\begin{eqnarray}
    \begin{aligned}
      &\nabla\times{\bf E}=-\partial_t\,\mu_0{\bf H}, \\
      &\nabla\times{\bf H}=\varepsilon_0\varepsilon_\infty \partial_t{\bf E}+{\bf J}, 
    \end{aligned}
  \label{eq:Maxwell}
\end{eqnarray}
where $\mu_0$ and $\varepsilon_0$ are the vacuum permeability and permittivity, respectively; $\varepsilon_\infty$ is the relative permittivity accounting for any dielectric local response. 
$\bf J$ is the current inside the electron gas.
This current is given by ${\bf J}=-en{\bf u}$ where $\bf u$ is the velocity of the electron gas, considered as a fluid, $e$ is the electron charge in absolute value, and $n$ is the electron density.
In this framework the velocity satisfies the following equations \cite{manfredi2001self,crouseilles2008quantum}:
\begin{eqnarray}
    \begin{aligned}
      &\partial_t n+{\nabla}\cdot(n{\bf u})=0\\
      &\partial_t{\bf u}+\gamma{\bf u}+{\bf u}\cdot{\bf \nabla}{\bf u}=\\
      &\quad=-\dfrac{e}{m}({\bf E}+{\bf u}\times\mu_0{\bf H})-\frac{1}{m}\nabla\frac{\delta G[n]}{\delta n}, 
    \end{aligned}
  \label{eq:u}
\end{eqnarray}
where $m$ is the electron mass and the energy functional $G[n]$ contains all the internal energy of the electronic system (here the operator $\frac{\delta}{\delta n}$ indicates the functional derivative with respect to the density $n$).
The last term in the second equation gives rise to a nonlocal contribution to the current $\bf J$.
Using  the previous equations,  we  obtain  the  following evolution equation for the current:
\begin{eqnarray}
    \begin{aligned}
      \partial_t{\bf J}+\gamma {\bf J}&-\frac{1}{e}\left(\frac{\bf J}{n}{\bf \nabla}\cdot{\bf J}+{\bf J}\cdot{\bf \nabla}\frac{\bf J}{n}\right)=\\
      &=\frac{ne^2}{m}{\bf E}-\frac{\mu_0e}{m}{\bf J}\times{\bf H}+\frac{en}{m}\nabla\frac{\delta G[n]}{\delta n},
    \end{aligned}
\end{eqnarray}
Neglecting higher order terms and taking $n({\bf r},t)=n_0({\bf r})+n_1({\bf r},t)$, and ${\bf E}({\bf r},t)={\bf E}_0({\bf r})+{\bf E}_1({\bf r},t)$, with $n_0$ being the equilibrium charge density and ${\bf E}_0$ the electrostatic field, we first obtain an equation for the dynamic part,
\begin{eqnarray}
    \begin{aligned}
    &(\partial_t+\gamma){\bf J}=\frac{e^2}{m}\left(n_0{\bf E}_1+n_1{\bf E}_0\right)+\\
    &\quad\quad+\frac{e}{m}\left[n_0\nabla\left(\frac{\delta G[n]}{\delta n}\right)_1+n_1\nabla\left(\frac{\delta G[n]}{\delta n}\right)_0\right],
    \end{aligned}
\end{eqnarray}
where $\left(\frac{\delta G[n]}{\delta n}\right)_0$ and $\left(\frac{\delta G[n]}{\delta n}\right)_1$ are the zero-th and first order terms of the potential, respectively. Second, we get an equation linking the static electron density to the static electric field,
\begin{equation}
\nabla\left(\frac{\delta G[n]}{\delta n}\right)_0+e{\bf E}_0=0, 
  \label{eq:zero-th}
\end{equation}
where ${\bf E}_0$ and $n_0$ can be found using Gauss's law:
\begin{equation}
\nabla\cdot\left(\varepsilon_\infty{\bf E}_0\right)=-\frac{e}{\varepsilon_0}(n_0-n^+), 
  \label{eq:gauss}
\end{equation}
 $n^+$ being the positive background density.

Finally, combining Eqs. (\ref{eq:Maxwell}) and assuming a harmonic field oscillating at $\omega$ we obtain:
\begin{eqnarray}
        &&\nabla\times\nabla\times{\bf E}_1-\varepsilon_\infty k_0^2{\bf E}_1=i\omega\mu_0{\bf J}\label{eq:wave}
        \\ 
    &&(\gamma-i\omega){\bf J}=\frac{e^2}{m}\left(n_0{\bf E}_1+n_1{\bf E}_0\right)+\nonumber
    \\
    &&\quad\quad+\frac{e}{m}\left[n_0\nabla\left(\frac{\delta G[n]}{\delta n}\right)_1+n_1\nabla\left(\frac{\delta G[n]}{\delta n}\right)_0\right],
    \label{eq:masterJ}
\end{eqnarray}
where $k_0=\omega/c$.

Combining Eqs.~(\ref{eq:zero-th}) and (\ref{eq:gauss}), we obtain:
\begin{eqnarray}
    \begin{aligned}
      &\nabla\left[\varepsilon_\infty \cdot \nabla \left(\frac{\delta G}{\delta n} \right)_{0}\right]-\frac{e^2}{\varepsilon_0}\left(n_0-n^+ \right)=0,\label{eq:scn0}
    \end{aligned}
\end{eqnarray}
Note also that using Eq.~(\ref{eq:zero-th}) we can simplify Eq.~(\ref{eq:masterJ}) as:

\begin{eqnarray}
    \begin{aligned}
      &(\gamma-i\omega){\bf J}=\frac{e^2}{m}n_0{\bf E}_1+\frac{en_0}{m}\nabla\left(\frac{\delta G[n]}{\delta n}\right)_1. \label{eq:QHT}
    \end{aligned}\label{eq:J}
\end{eqnarray}

\subsection{LRA in the presence of spill-out}
We first derive the equations corresponding to the LRA \cite{Esteban2012NatCommun,SkjolstrupPhys.Rev.B.,Skjolstrup2019_PRB}. They have the merit to allow a better grasp of the physics at play. The main assumption here is to neglect the pressure term, i.e. $G[n]\simeq 0$, making the description purely local.

The impact of nonlocality alone in the case of vanishing gaps (and hard-wall boundaries) has already been investigated in Ref. \cite{Raza2013Phys.Rev.B.}, where it has been shown that the effective index of the gap-plasmon increases although staying finite (contrarily to the diverging behavior of the LRA case without spill-out \cite{Bozhevolnyi2008Opt.Exp., Smith2015Nanoscale}). Yet, it does not tend to the expected value of the bulk metal refractive index either.

Starting from Eq. \eqref{eq:J} we get 
\begin{eqnarray}
    \begin{aligned}
      &{\bf J}=\frac{1}{\gamma-i\omega}\dfrac{e^2}{m}n_0{\bf E}_1.
    \end{aligned}
\end{eqnarray}
The current can always be incorporated into Eqs. (\ref{eq:Maxwell}) as an effective polarization $\textrm{P}_f$, by simply writing  $\partial_t \textrm{P}_f =\textrm{J}$. This finally allows to introduce a local susceptibility by writing, just as for a Drude model, that 
\begin{equation}
    \mathbf{P}_f = -\varepsilon_0 \frac{\frac{e^2n_0(\mathbf{r})}{\varepsilon_0 m}}{\omega^2 - i\gamma \omega} {\mathbf E}_1 =\varepsilon_0 \chi_f {\mathbf E}_1,
\end{equation}
except that the susceptibility is proportional to the electron density and thus depends on the position here.

Now if we consider the propagation of a guided mode in the $z$ direction, and assume that all fields are invariant in the $x$ direction, Maxwell's equations in the $p$-polarization reduce to:
\begin{align}
&\partial_y E_z -\partial_z E_y = i\omega\mu_0 H_x, \\
&\partial_z H_x =-i\omega \varepsilon_0 \varepsilon(y) E_y, \\
-&\partial_y H_x = -i\omega \varepsilon_0 \varepsilon(y) E_z,
\end{align}
where $\varepsilon(y) = \varepsilon_\infty + \chi_f(y)$ is the local permittivity.

Combining the previous equations to get rid of $H_x$ we obtain,
\begin{align}
  \partial_y \partial_z E_z -\partial^2_z E_y &= i\omega \mu_0 \partial_z H_x =k_0^2 \varepsilon(y)  E_y\\
  \partial_y^2 E_z -\partial_y \partial_z E_y &= i\omega \mu_0 \partial_y H_x = -k_0^2 \varepsilon(y) E_z
\end{align}

Since we are looking for a guided mode we take $\partial_z = i k_z$ so that we obtain
\begin{equation}
    (k_z^2 - \varepsilon(y) k_0^2 ) E_y = -ik_z \partial_y E_z,
\end{equation}
and finally manage to keep only the $E_z$ field to write the wave equation under the form
\begin{equation}
  \partial_y^2 E_z + k_0^2\, \varepsilon(y)\, E_z + \partial_y \left(
  \frac{k_z^2}{\varepsilon(y)\, k_0^2-k_z^2} \partial_y E_z 
  \right)=0.\label{symp_mod}
\end{equation}

We underline that such an approach is not self-consistent, as it requires to choose arbitrarily the electron density. 
The choice of electron density put aside, our local model with spill-out is identical to the model used in \cite{SkjolstrupPhys.Rev.B.} and yields perfectly similar results. In Ref. \citenum{SkjolstrupPhys.Rev.B.}, the authors compute $n_0$ using a quantum mechanical approach whereas we use a simple model density to approximate the electron spill-out (details are given in the next section).

We have then converted this equation into a numerical problem using a finite difference scheme, in order to look for values of $k_z$ in the complex plan which allows to make the determinant of the finite difference matrix vanish. As such a problem is extremely unstable numerically, obviously because of the different scales involved (the gap being much smaller than the skin depth), we have used the exponentially decaying analytic solution inside the metal. This solution is known when $\varepsilon(y)$ is uniform and corresponds to the classical solution of Maxwell's equations considering Drude's model for the metal. Connecting the finite difference problem with the analytic solution allows to reduce the size of the problem and to improve the stability, finally allowing to compute the wave vector of the gap-plasmon, even for vanishing gaps. We compared it with a finite element based approach, and both are in almost perfect agreement (see Appendix).

\subsection{Self-consistent quantum hydrodynamic theory}
Equation~(\ref{eq:scn0}) with Eqs.~(\ref{eq:wave}) and (\ref{eq:QHT}) constitute the basis of the QHT \cite{Ciraci2016Phys.Rev.B,Toscano2015NatCommun} linear response and allow to calculate the detailed electron gas dynamic at the metal surface.
In this work, we employ the following approximation:
\begin{align}
G[n]= T_{\textrm{TF}}[n]+\lambda_{\rm vW} T_{\textrm{vW}}[n, \nabla n]+E_{\textrm{XC}}[n] \label{eq:G[n]}
\end{align}
where $T_{\rm TF}$ is the TF kinetic energy, $T_{\rm vW}$ is the gradient-dependent correction, namely the von Weizs\"acker term, to the TF functional, which allows to take into account effects depending on the gradient of the electron density $n$ and thus should not be neglected when spatially dependent densities are considered.
$E_{\rm XC}$ is the exchange-correlation energy functional.
The parameter $\lambda_{\rm vW}$ in front of the von Weizs\"acker term can be related to the decay of the charge density from a metal surface, and hence controls the amount of the electron spill-out and tunneling. It has been shown in Ref.~\citenum{Ciraci2016Phys.Rev.B} that for QHT $\lambda_{\rm vW}$ is inversely proportional to the decay of the charge density and can be related to the ``equilibrium" and ``induced" electron spill-out. 
In general, $\lambda_{\rm vW}$ is a free parameter of the kinetic functional that can be tuned depending on what the figure of merit is (\textit{i.e.}, plasmon energy, equilibrium density asymptotic decay, tunneling, etc).
Most often, in the literature a value in the $[1/9, 1]$\textendash range is used for $\lambda_{\rm vW}$; in which $\lambda_{\rm vW}=1/9$ corresponds to a lower electron spill-out and a faster decay of $n_0$ from a metal surface, whereas $\lambda_{\rm vW}=1$ gives a higher spill-out and a slower decay of the charge density. In general, using $\lambda_{\rm vW}=1/9$ approximates well the Lindhard function for small $k$\textendash vectors while $\lambda_{\rm vW}=1$ gives a good estimation for large $k$\textendash vectors \cite{Wang2000}.
In fact, the von Weizs\"acker correction in Eq.~\eqref{eq:G[n]} is the first-order term in the expansion of the kinetic energy \cite{Manfredi2012NJP} and to construct a more generally valid functional one would need to consider  higher-order terms (\textit{i.e.}, Laplacian dependence), which would introduce more free parameters \cite{Baghramyan2020}.\\
\indent The explicit expressions of the energy functional $G[n]$ and its functional derivative, i.e. $\left( \frac{\delta G}{\delta n} \right)_1$, can be found in Ref.~\citenum{Ciraci2016Phys.Rev.B}. We further note that Landau damping is another important factor to consider and it can also be incorporated in Eq.~\eqref{eq:J}, as was done in Ref.~\citenum{Ciraci2017Phys.Rev.B} in the context of QHT. We expect that the Landau damping will result in increased damping inversely proportional to the gap size. Since the main objective of the present work is to analyze the impact of nonlocality and electron spill-out on the gap plasmons, we neglect Landau damping. It is informative to note that considering a spatially-constant $n_0=n_b$, $n_b$ being the charge density of the bulk metal, and $G[n]=0$ in Eq.~(\ref{eq:G[n]}) leads to the usual Drude-type relation (LRA) for the polarization, and using $T_{\textrm{vW}}=E_{\textrm{xc}}=0$ returns $G[n]=T_{\rm TF}$, the standard hydrodynamic theory in the Thomas-Fermi approximation (TFHT). 

In contrast to the conventional theories in plasmonics, QHT can efficiently take into account the nonlocal, electron spill-out and tunneling effects, and can provide the details of microscopic distribution of the fields which usually cannot be described by conventional approaches. The QHT is highly dependent on the spatially varying equilibrium charge density $n_0$ and, consequently, approximation of the optical response of a nanoplasmonic system relies strongly on the description of $n_0$. 
Over the past few years, QHT has been applied to probe linear plasmonic properties of a variety of individual nanoparticles as well as nearly-touching nanostructures (i.e., in the tunneling regime) \cite{Ciraci2016Phys.Rev.B, Khalid2018Opt.Express, Khalid2019Photonics, Ciraci2019NanoPhotonics} and a fairly good agreement between the QHT and TD-DFT calculations has been reported \cite{Khalid2018Opt.Express}. Very recently, electron spill-out effects in singular metasurfaces have been analyzed using QHT, showing that the spill-out effectively blunts the sharp singularities \cite{Yang2021Photonics}. QHT has also been applied to study nonlinear optical properties of thick metal films and it has been shown that there exist spill-out induced resonances which can be exploited to achieve very large second-harmonic generation efficiencies \cite{Khalid2020commphys}. These resonances cannot be excited in the LRA and TFHT, as electron spill-out is overlooked in these approximations. Therefore, it becomes important to take into account the nonlocal and quantum effects when dealing with plasmonic structure with sub-nanometer geometrical features.
Despite the fact that QHT gives a very good prediction of the plasmon resonances by adding a $\nabla n-$dependent correction to the kinetic energy, however, it also suffers some limitations.
For example, it predicts unexpected modes between the surface and bulk plasmon frequencies lying in the far tail region of the exponentially decaying charge density.  
Very recently, a development in the QHT theory based on the Laplacian-level kinetic energy functionals has been presented \cite{Baghramyan2020}. 
The introduction of Laplacian-dependent charge density results in more robust numerical solutions but its implementation becomes more complex.
Since in the present study we only consider very small gaps, the tail of the charge density never extends to the problematic region.
Therefore, we expect that the QHT approach in the limit of the von Weizs\"acker approximation used herein gives us results with the same level of accuracy and with the advantage of avoiding additional numerical complexity, arising from the Laplacian-level functionals.
We adopt the jellium approximation \cite{Brack1993Rev.Mod.Phys.} in implementation of the QHT, which assumes that the electrons in a metal are confined by a constant positive background charge $n^+=(\frac{4}{3}\pi r_s^3)^{-1}$, where $r_s$ indicates the Wigner-Seitz radius ($r_s=4$ for Na and $r_s=3$ for Ag).

In order to compute the gap-plasmon mode and its dispersion relation we first compute the space-dependent equilibrium charge density using Eq.~(\ref{eq:scn0}) and then solve Eqs.~(\ref{eq:wave}) and (\ref{eq:QHT}) assuming a solution of the type $\mathbf{E}(\mathbf{r})=\tilde{\mathbf{E}}(y)\,e^{ik_{\rm gp}z}$ (with similar expressions for $\mathbf{J}$), where $k_{\rm gp}$ is the propagation  constant  of  the  gap  plasmon mode.
For a given angular frequency $\omega$, we solve this system of equations using the finite-element method to compute the associated propagation constant $k_{\rm gp}(\omega)$, as well as the mode $\tilde{\mathbf{E}}(y)$.

\section{Results}\label{sec:results}

\begin{figure}
	\centering
	\includegraphics[width=1\linewidth]{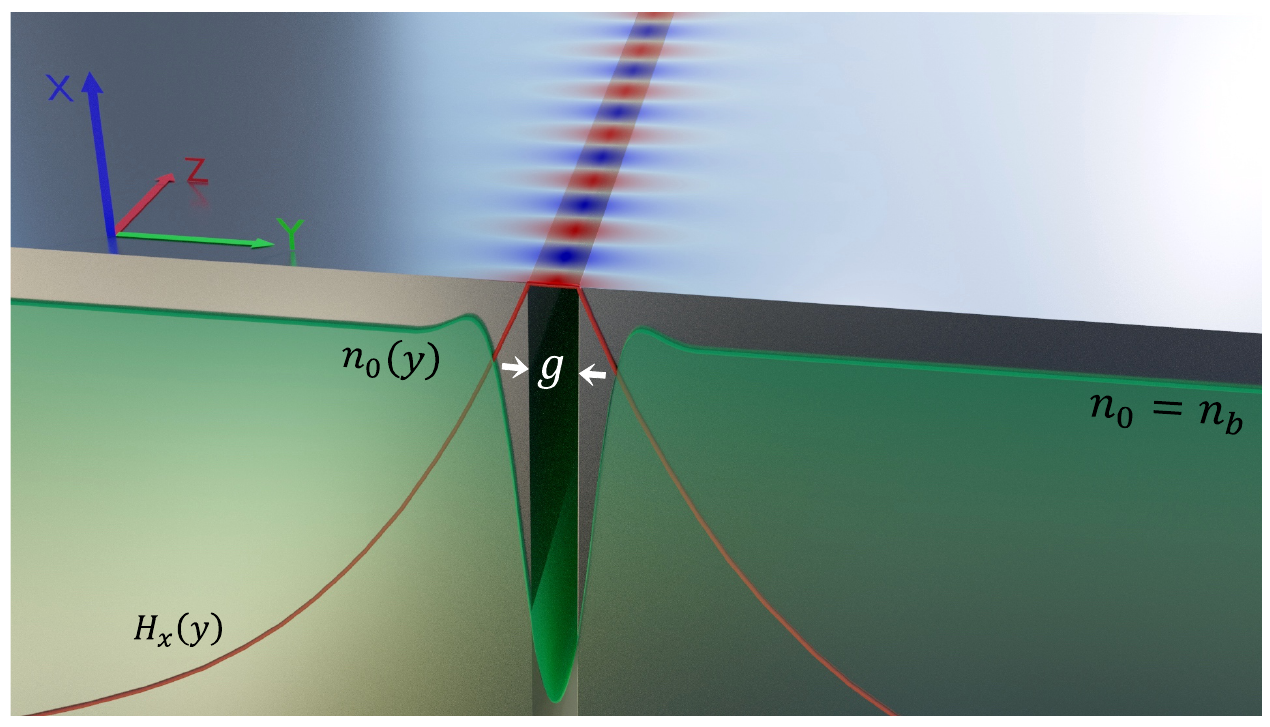}
	\caption{Visual representation of overlapping of the equilibrium charge densities $n_0(y)$ due to electron spill-out in the dielectric gap $g$ between two semi-infinite bulk metals. Deep inside the bulk metal, $n_0$ is equal to the constant density of the bulk metal $n_b$. The magnetic mode profile $H_x(y)$ of a $p-$polarized light propagating along the $z-$direction is also shown. The structure is invariant along the $x-$ and $z-$directions.}
	\label{fig:n0_metal_feat}
\end{figure}

Let us consider a MIM configuration as schematically shown in Fig.~\ref{fig:n0_metal_feat}, that is, a dielectric gap $g$ sandwiched between two semi-infinite metallic regions. For simplicity we use air as the dielectric. We examine the impact of nonlocality and the electron spill-out on the gap-plasmon, the unique mode guided in $p$-polarization in the structure and propagating along the $z$ direction. More precisely, we compute the effective index $n_{\rm eff} = \frac{k_z}{k_0}$ of the gap-plasmon according to the different models described above and compare the results, in particular in the gap size range ($<$ 1 nm) where the charge densities from the two surfaces significantly overlap, as schematically shown in Fig.~\ref{fig:n0_metal_feat}. We begin with the LRA with spill-out for simple Drude metals, i.e., sodium (Na), which we compare with the LRA and TFHT approximations without spill-out as well as with the QHT. Finally, we consider the case of silver (Ag) in the QHT framework and compare to the commonly used LRA and TFHT without spill-out.

\subsection{LRA with spill-out}
We begin with the simplified case presented above in which the equilibrium electron density $n_0(\mathbf{r})$ can be described by an exponentially decaying profile with the following analytical function,
\begin{equation}
    n_0(y)= n_b\left[\frac{1}{1+e^{\kappa \left(y+\frac{g}{2}\right)}}+\frac{1}{1+e^{-\kappa\left(y-\frac{g}{2}\right)}}\right] \label{eq:an_n0}
\end{equation}
where the parameter $\kappa$ defines the spill-out or the decay of the charge density from the metal surface. In what follows, we set $\kappa\simeq(0.7 a_0)^{-1}$, with $a_0$ being the Bohr radius, which we get by fitting of the analytical function with the decay of the self-consistent charge density (see Eq.~(\ref{eq:scn0})), considering $\lambda_{\rm vW}=1/9$.
We take $r_s=4$, $\varepsilon_\infty=1$ and $\hbar \gamma=0.16$~eV, values that correspond to Na. Although it is impossible to work with Na in practical applications, due to its high reactivity, it presents an ideal Drude response as the impact of interband transitions can be omitted, thus offering a convenient theoretical study platform.

\begin{figure}
\centering
\includegraphics[width=0.9\linewidth]{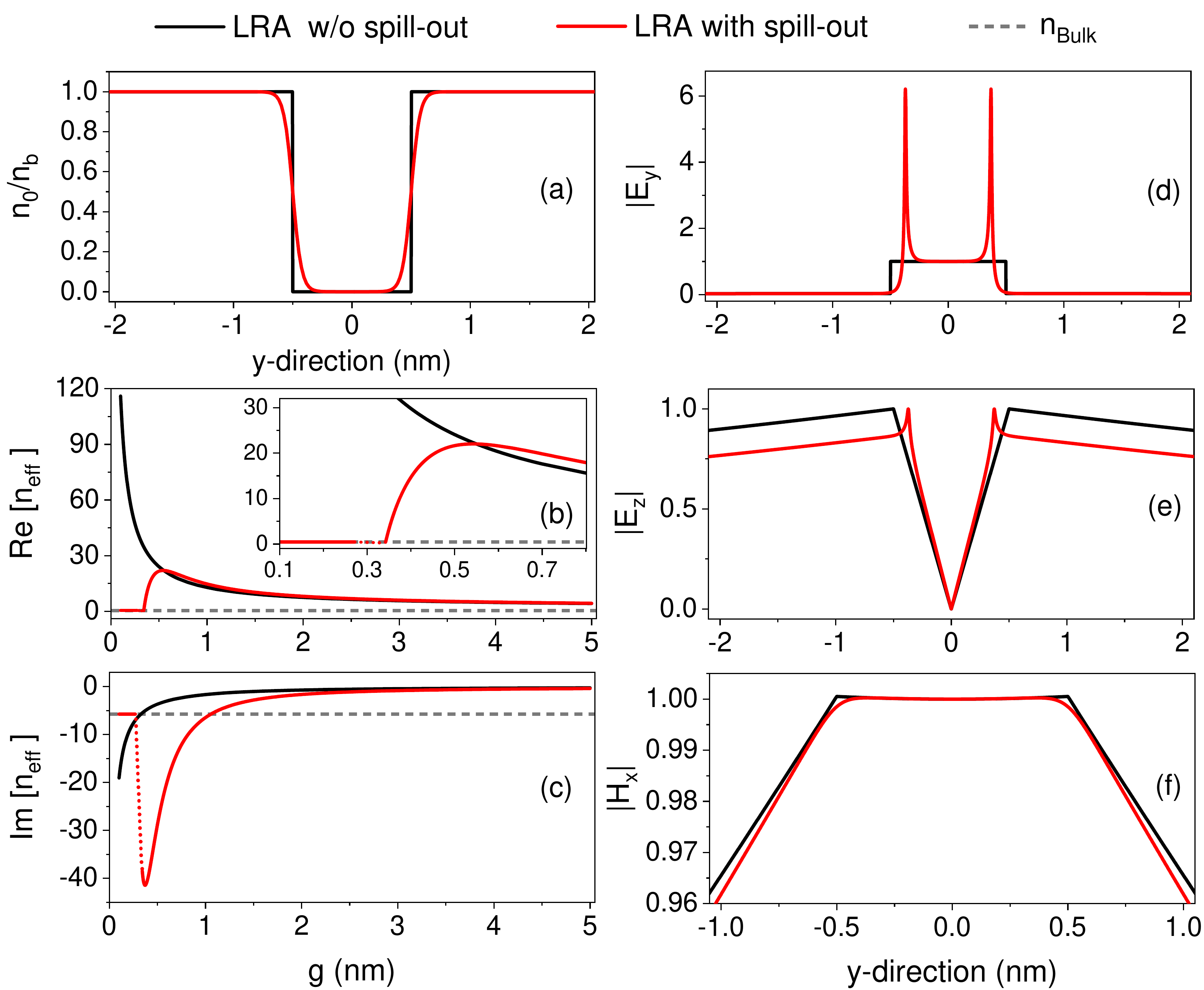}
\caption{A comparison between the local response approximation (LRA) with and without electron spill-out from the metal surfaces. (a) Equilibrium electron density $n_0$ normalized with the bulk density $n_\mathrm{b}$ for $g=1~$nm. The real (b) and (c) imaginary parts of the effective refractive index $n_\mathrm{eff}$ plotted at $E=1~$eV. The inset in (b) depicts a zoom-in of the region where the spill-out has a significant impact on the mode index. The horizontal dashed-grey line marks the refractive index of the bulk metal, $n_{\rm Bulk}$. The dotted-red lines represent the region where it becomes hard to find a proper solution due to numerical artifacts. (d)-(f) present a comparison between the electric and magnetic field components considering $g=1~$nm and $E=1~$nm. The plots in the right column are normalized with the field value at the center of the gap, except in (e) where field is zero at the center of the gap and the maximum value of the field is considered for normalization.} \label{fig:local_LRAso}
\end{figure}

In Fig.~\ref{fig:local_LRAso} we compare the results of our simplified LRA with spill-out against the commonly used LRA without any spill-out. The input equilibrium charge density profiles used in the both approaches are plotted in Fig.~\ref{fig:local_LRAso}(a) for $g=1$~nm.
In Fig.~\ref{fig:local_LRAso}(b) and (c), we compare the behavior of the real and imaginary parts of the effective mode index. As expected, for large gaps ($g>1~$nm) spill-out can be neglected, the LRA with and without electron spill-out agree very well. The spill-out has a noticeable impact on the real part of the effective index for gaps that are typically around 1 nm or less: it increases the effective index, signaling a plasmonic slow down of the mode which is larger than without any form of spill-out. This is understandable, as the Poynting vector inside a plasma is opposite to the propagation direction, leading to a plasmonic drag \cite{ajib2019energy}, to a lower group velocity and finally to a larger wave vector. 

When the gap is as small as 0.5 nm it begins to ``close", electromagnetically speaking: the electron density inside the gap becomes so large that the propagation of the gap-plasmon is hindered. The real part of the effective index decreases quickly, while its imaginary part is much higher when the spill-out is taken into account. It is important to note that in the limit of vanishing gaps, where the spill-out is expected to virtually close the gap, both real and imaginary parts of $n_{\rm eff}$ should ideally converge to the index of the bulk metal. In Fig.~\ref{fig:local_LRAso}(c), we notice that when $g < 0.35~$ nm, i.e., the gap distances where the gap is expected to vanish completely, the imaginary part of $n_{\rm eff}$ tends to bend back. However, we have found to be very hard to obtain a proper solution in this region. Indeed the dotted line as shown in Figs.~\ref{fig:local_LRAso}(b) and (c) represents a region where we could not find a solution due to numerical artifacts. When $g<0.28~$nm, we find $n_{\rm eff}=n_{\rm Bulk}$, where $n_{\rm Bulk}$ being the refractive index of the bulk metal. These results are in agreement with the ones reported in \cite{SkjolstrupPhys.Rev.B.}.

We attribute the large imaginary part of the effective index to peaks in the electric field that can be seen when the electron density reaches a value so that $\chi_f \simeq 0$, meaning a vanishing local permittivity $\varepsilon(y)$. The components of the electric field in the transverse and longitudinal directions with respect to mode propagation are shown in Figs.~\ref{fig:local_LRAso}(d) and (e), respectively. The extremely narrow peaks the electric field exhibits have already been seen in a similar context \cite{SkjolstrupPhys.Rev.B.} with a different electron gas density. This means that they are fundamentally linked to the assumption that a local permittivity is able to describe the spill-out accurately. A clue that nonlocality may play an important role here is that $\varepsilon=0$ is the condition for which volume plasmons are expected to be supported inside the electron gas.

The peaks in the electric field drastically increase the losses experienced by the gap plasmon, since the local absorption by the electron gas is proportional to  $\mathbf{E}_1\cdot\mathbf{J}^*$, where $^*$ denotes the complex conjugate \cite{sakat2021}. This explains the larger imaginary part. We underline that, on the contrary, the magnetic field remains smooth (see Fig.~\ref{fig:local_LRAso}(f)).

\subsection{QHT for simple metals}

In the framework of QHT, we compute the self-consistent ground-state density of a symmetric MIM waveguide from Eq.~(\ref{eq:scn0}). First we consider that the guiding structure is composed of an air gap sandwiched between two semi-infinite Na metals, to better grasp the difference with the simplified spill-out model. 

\begin{figure}
	\centering
	\includegraphics[width=0.9\linewidth]{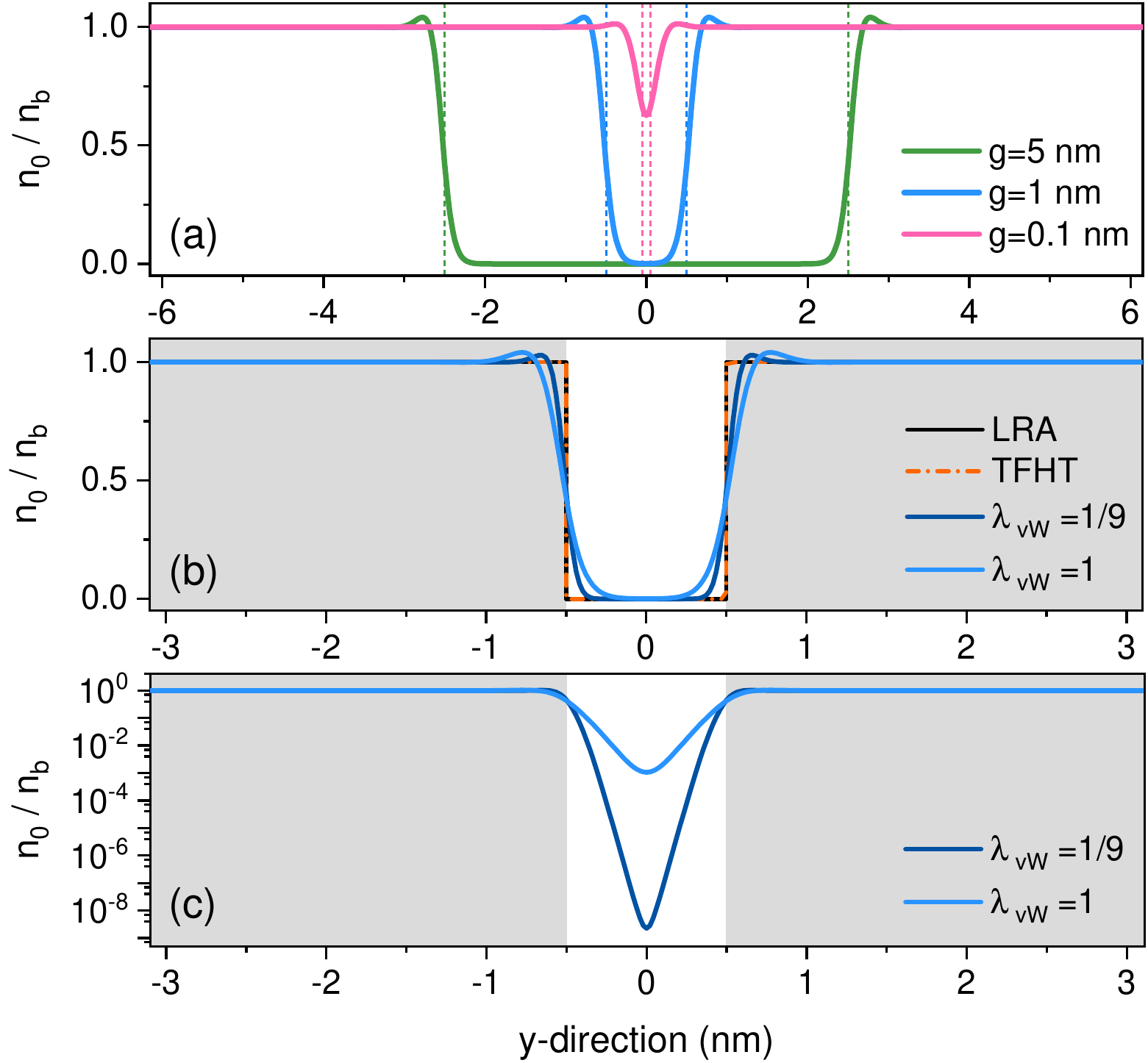}
	\caption{Equilibrium charge density $n_0$ calculated self-consistently for an air gap sandwiched between the Na metals. (a) For different gap size ($g$). The vertical dashed lines represent the metal boundaries for the corresponding gaps. (b) A comparison between the LRA, TFHT and QHT for the spill-out parameter $\lambda_{\rm vW}=1/9$ and $\lambda_{\rm vW}=1$, considering $g=1$~nm and (c) the decay of the charge density at $\lambda_{\rm vW}=1/9$ and $1$. The results are normalized with the bulk density $n_b$ and the shaded grey area represents the metal region.}
	\label{fig:Na_n0sc}
\end{figure}

The self-consistent ground-state density $n_0(\mathbf{r})$ for a Na-air-Na configuration is shown in Fig.~\ref{fig:Na_n0sc}~(a) considering different gaps. 
The metal-dielectric interfaces for each gap size are indicated by the dashed lines.
It can be seen that when the inter-distance of the metal surfaces is sufficiently large, there is no overlap of the charge densities.
In this scenario, the system can be well described using a classical formulation.
However, when the gap spacing between the metal boundaries decreases, the charge density profiles start to overlap and the electrons from the two surfaces start to tunnel across the gap.
For sufficiently small gaps the charge transfer between the metals is large enough to close the gap optically speaking, despite the fact that the metal ion surfaces do not touch physically.
Figure~\ref{fig:Na_n0sc}~(b) presents a comparison of $n_0$ between the LRA, TFHT and QHT for gap size $g=1$~nm.
In the LRA and TFHT approximations, we assume a constant $n_0=n_b$ in the metal and zero outside.
In the QHT case, the space-dependent $n_0(\mathbf{r})$ is plotted for both $\lambda_{\rm vW}=1/9$ and $1$.
The parameter $\lambda_{\rm vW}=1/9$ represents a lower spill-out and a faster decay of the charge density, whereas $\lambda_{\rm vW}=1$ gives a higher spill-out and a slower decay, as shown in Fig.~\ref{fig:Na_n0sc}~(c).        
\noindent 

\begin{figure}
	\centering
	\includegraphics[width=0.9\linewidth]{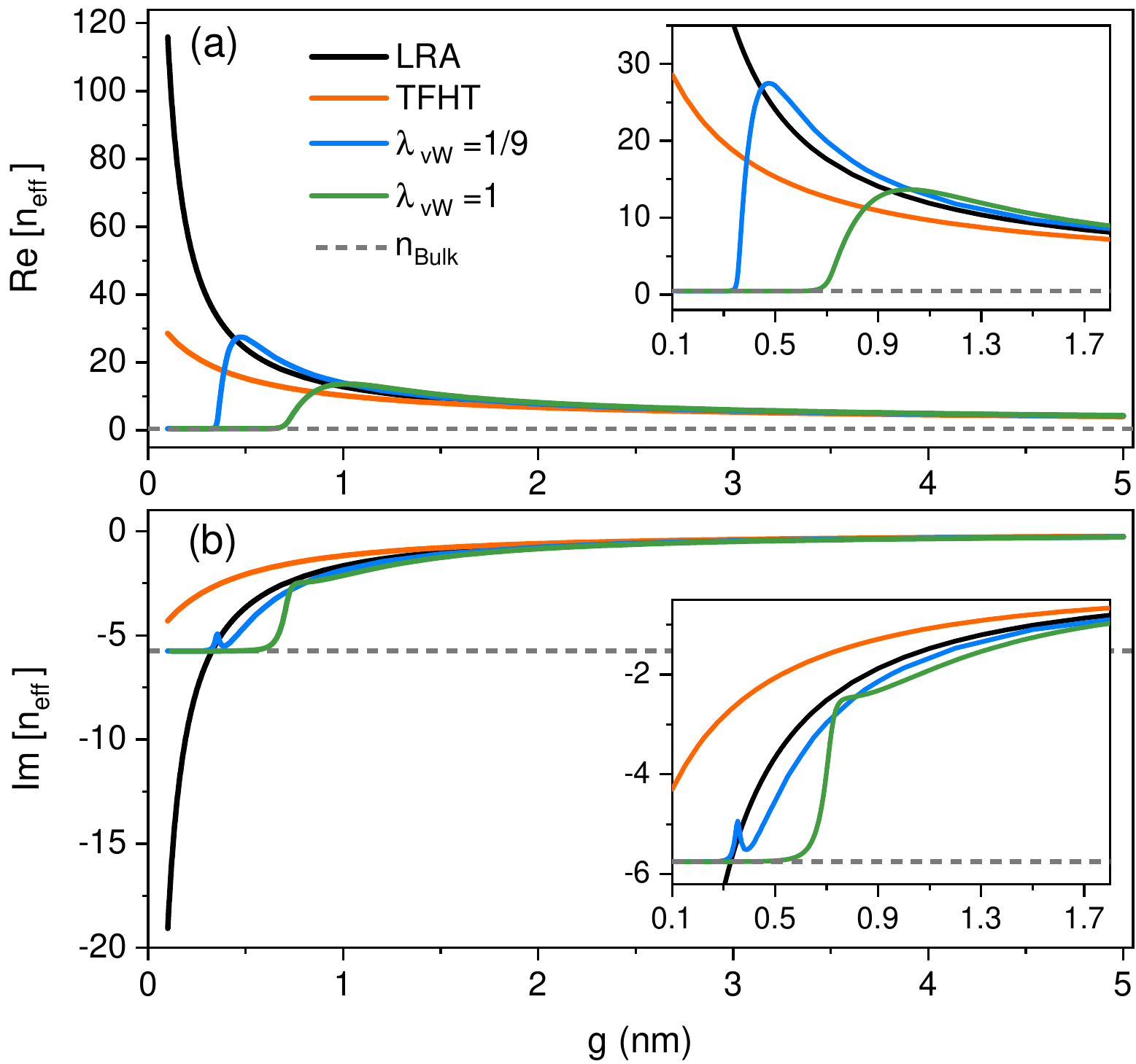}
	\caption{Effective refractive index $n_{\rm eff}$ of Na-air-Na configuration as a function of gap distance $g$. A comparison between the LRA, TFHT and QHT is given. (a) Real and (b) imaginary parts of $n_{\rm eff}$, plotted at $E=1$~eV. The horizontal dashed-grey lines represent the refractive index of the bulk metal.}
	\label{fig:Na_gapSweep}
\end{figure}

In Fig.~\ref{fig:Na_gapSweep}, we plot the effective refractive index as a function of gap distance $g$.
The results computed within the QHT, considering the self-consistent space-dependent charge density, are compared against the LRA and TFHT without spill-out.
It is interesting to note that for extremely small gap separation, LRA predicts very large values of both real and imaginary parts of $n_{\rm eff}$, whereas the TFHT gives relatively smaller values. Nonetheless, the trend is quite similar, {\em i.e.} $n_{\rm eff}$ increases monotonically as the gap decreases.
In contrast, in the QHT approximation, the $n_{\rm eff}$ first increases with the decreasing gap size and then rapidly decrease after a certain gap distance, getting a value of the refractive index of the bulk metal.
This is due to fact that at a certain gap size, a reasonable charge transfer through the gap occurs between the metal surfaces due to the overlapping of the charge densities and a virtual fade-out of the physical gap appears when the spacing between the metal boundaries is further decreased.
For sufficiently small gaps, the mode virtually vanishes due to the electron tunneling and both the real and imaginary parts of the $n_{\rm eff}$ converges to the real and imaginary parts of the refractive index of the bulk metal, as shown in the inset of Fig.~\ref{fig:Na_gapSweep}(a) and (b), respectively. The expected limit for the effective index of the gap-plasmon is in that case actually retrieved.
In the QHT case, results for both $\lambda_{\rm vW}=1/9$ and $\lambda_{\rm vW}=1$ are presented. Since $\lambda_{\rm vW}=1/9$ entails the lower spill-out, the gap virtually vanishes when $g<0.3$~nm, however, the gap disappears even at higher separation ($g<0.45$~nm) for $\lambda_{\rm vW}=1$, as it implies higher electron spill-out.
\begin{figure}
\centering
\includegraphics[width=0.9\linewidth]{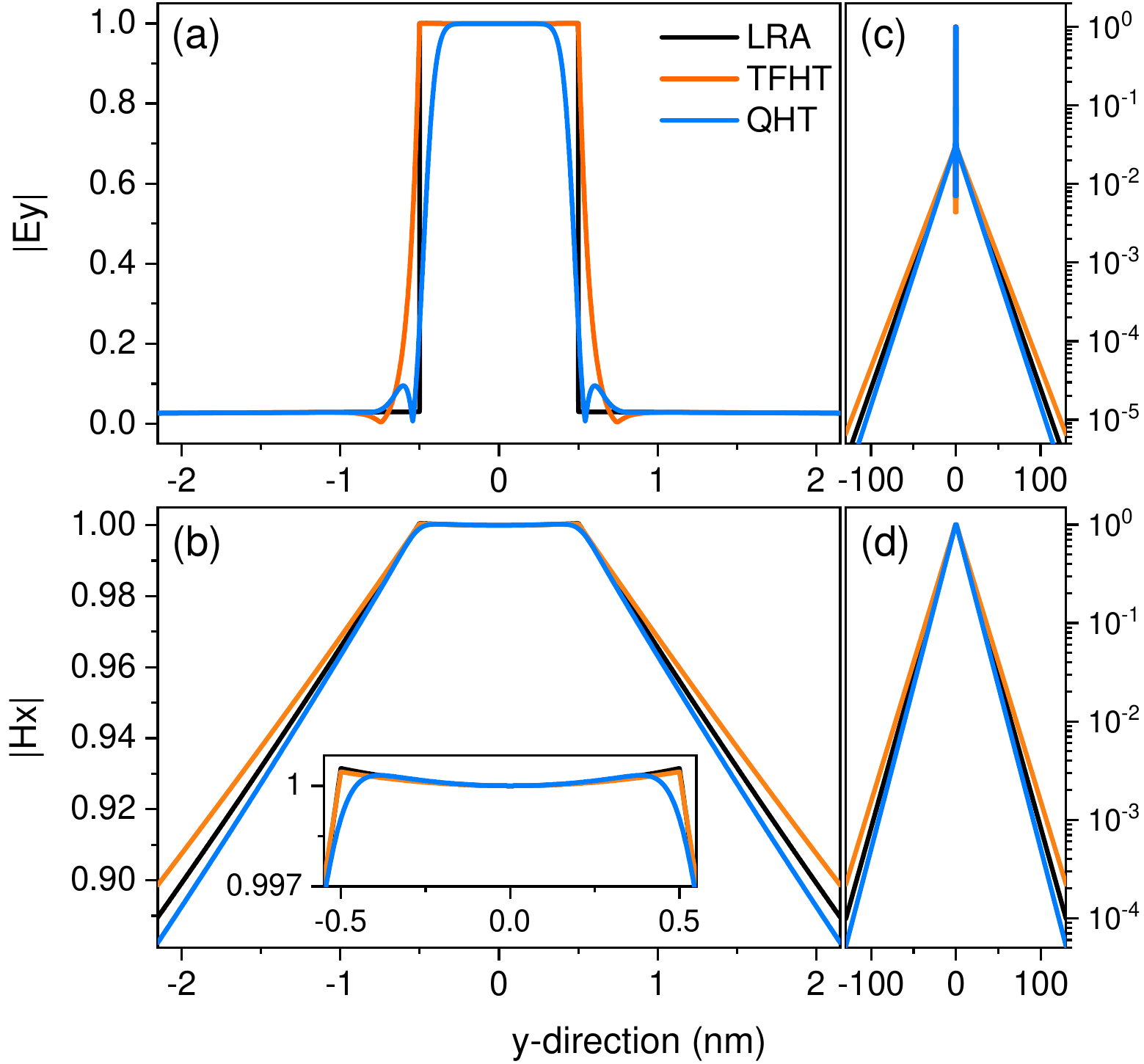}
\caption{(a) Normalized electric $|E_y|$ and (b) magnetic field $|H_x|$ field components plotted along the y-direction, i.e., in the normal direction to the interfaces, plotted uder different approaches. The inset of (b) is a close-up view of the $|H_x|$ at the interfaces. The plots are normalized by the field value at the center of the gap. (c) and (d) show decay of the fields at longer distances inside the metal. In these plots, we have considered $g=1$~nm, $E=1$~eV and $\lambda_{\rm vW}=1/9$ in the QHT case.} \label{fig:Fields_QHT}
\end{figure}

Now, to show the influence of the electron spill-out on the mode profile, we plot here the electric $|E_y|$ and magnetic $|H_x|$ fields in the orthogonal direction to the metal-air interfaces under different approximations for $g=1$~nm and $E=1~$eV, as shown in Fig.~\ref{fig:Fields_QHT}.
Both in the LRA and TFHT, $|E_y|$ is constant throughout the gap and then exponentially decays in the metal, as can be seen in Figs.~\ref{fig:Fields_QHT}~(a) and (c).
QHT, on the other hand, predicts that $|E_y|$ has a maximum value at the center of the gap and it decreases as moving away from the center. The effect of the electron pressure in the metal in the TFHT can also be observed on the field profile. 
The $|H_x|$, as plotted in Fig.~\ref{fig:Fields_QHT}~(b), remains quite similar in the gap in all approximations expect near the interfaces where QHT predict a bit lower value. The decay of fields is given in Fig.~\ref{fig:Fields_QHT}~(c) and (d), which show that the fields decay faster in QHT and slower in TFHT with respect to the LRA fields, essentially because the wavevector along $z$ and thus the decay along the $y$ axis is slightly different.

\begin{figure}
\centering
\includegraphics[width=0.9\linewidth]{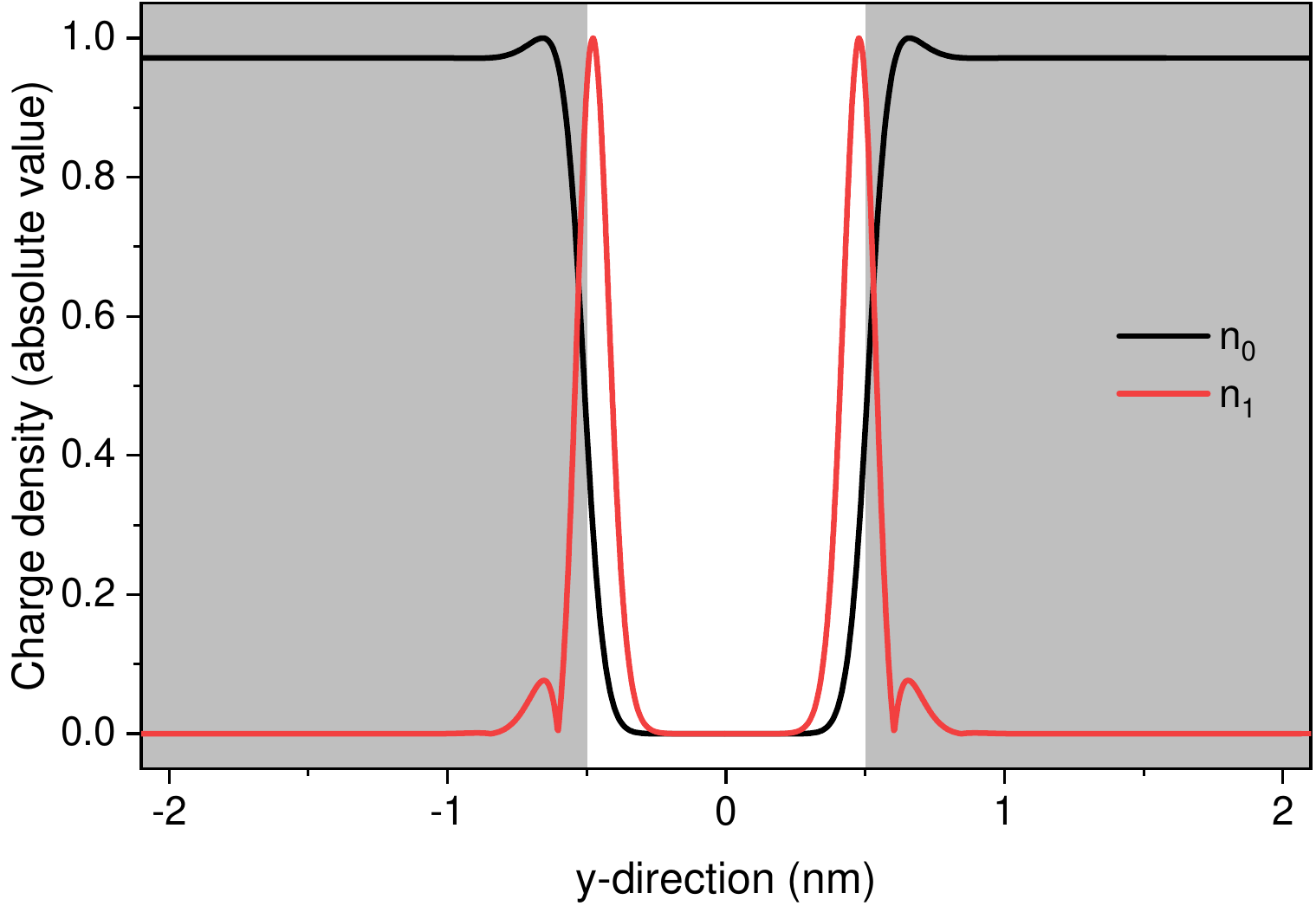}
\caption{Normalized equilibrium charge density $n_0$ and induced charge density $n_1$ computed using the QHT with $\lambda_{\rm vW}=1/9$. The other parameters are the same as used in Fig.~\ref{fig:Fields_QHT}. The shaded grey area represents the metal regions.} \label{fig:n1_Na}
\end{figure}

Now, we analyze the impact of electron spill-out and tunneling on the induced charge density. We, in fact, consider the spill-out and tunneling in both the equilibrium charge density and in the linear response, as the modification of the equilibrium density gives a more enhanced overlap of the induced charge densities. The induced charge density indeed is the equivalent of the transition density in the TD-DFT, and, in the limit of the kinetic energy functional approximation, represents the sum of all possible electron transitions, including those describing the tunneling. As an example, we plot in Fig.~\ref{fig:n1_Na} the induced charge density $n_1$ along with the equilibrium charge density $n_0$ in the same scenario as considered in Fig.~\ref{fig:Fields_QHT}. The peak of the induced charge density from each interface lies outside the jellium edge and the tails of the densities go more into the dielectric medium and consequently showing more overlap.

\begin{figure}
\centering
\includegraphics[width=0.9\linewidth]{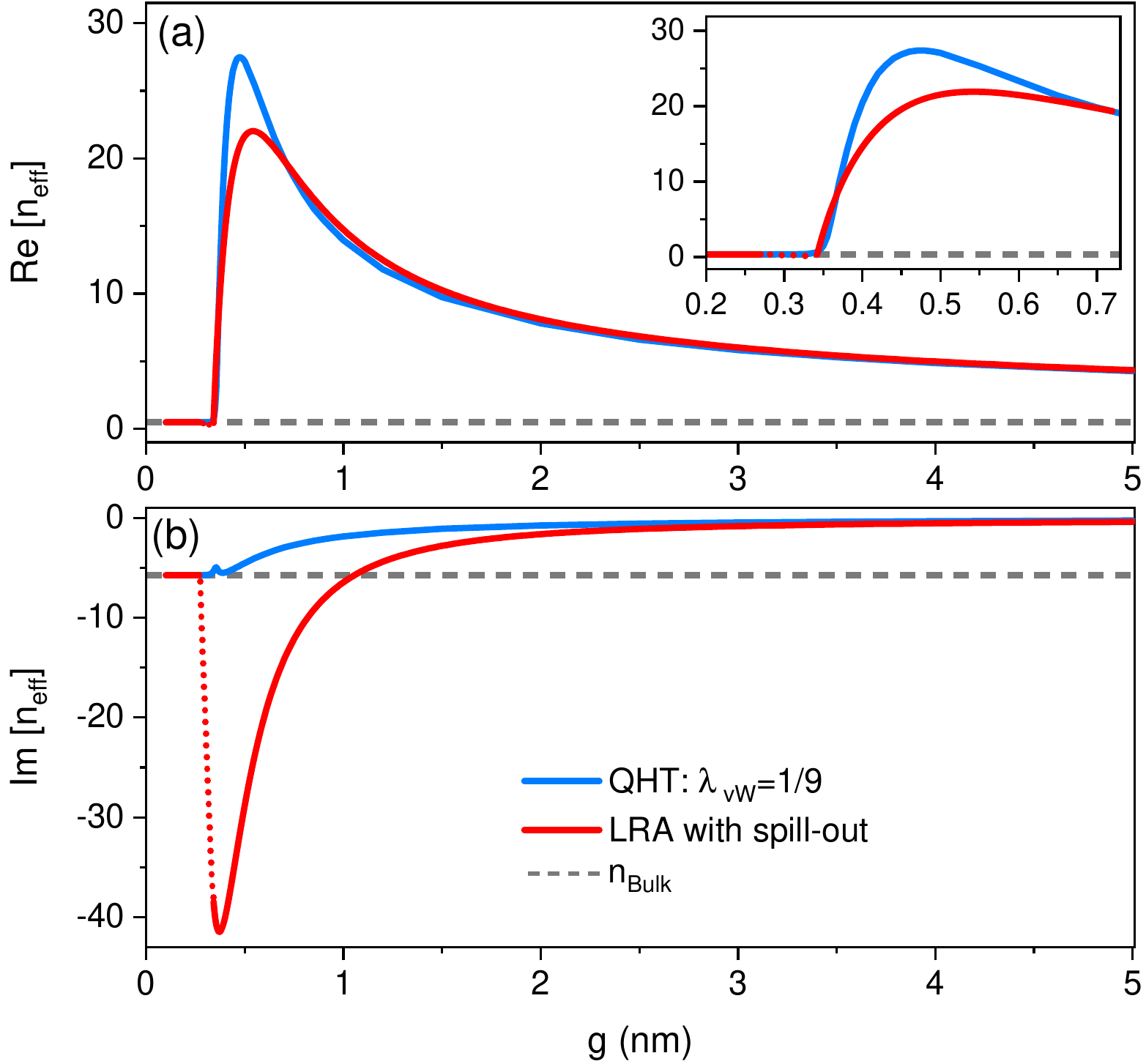}
\caption{(a) Real and (b) imaginary parts of $n_\mathrm{eff}$ within the LRA with electron spill-out and full self-consistent QHT for $\lambda_{\rm vW}=1/9$. The inset shows a zoom-in of the effective mode index in the tunneling regime. The dotted-red lines represent the gap sizes where it becomes hard to find a proper solution due to numerical artifacts, as was discussed in the caption of Figs.~\ref{fig:local_LRAso}(b) and (c).} \label{fig:QHT_vs_LRAso}
\end{figure}

Finally, it is interesting to compare the QHT and the simplified LRA with spill-out, as shown in Fig.~\ref{fig:QHT_vs_LRAso}. The real part of $n_{\rm{eff}}$ in both cases accord well until the gap closes (Fig.~\ref{fig:QHT_vs_LRAso}~(a)) whereas, as described previously, the imaginary part is much higher in the simplified spill-out model (Fig.~\ref{fig:QHT_vs_LRAso}~(b)). This really means that nonlocality avoids the existence of the peaks in the electric field when the electron density reaches a particular value, making the imaginary part much closer to the prediction of the LRA without spill-out.

\subsection{QHT for noble metals}

\begin{figure}
	\centering
	\includegraphics[width=0.9\linewidth]{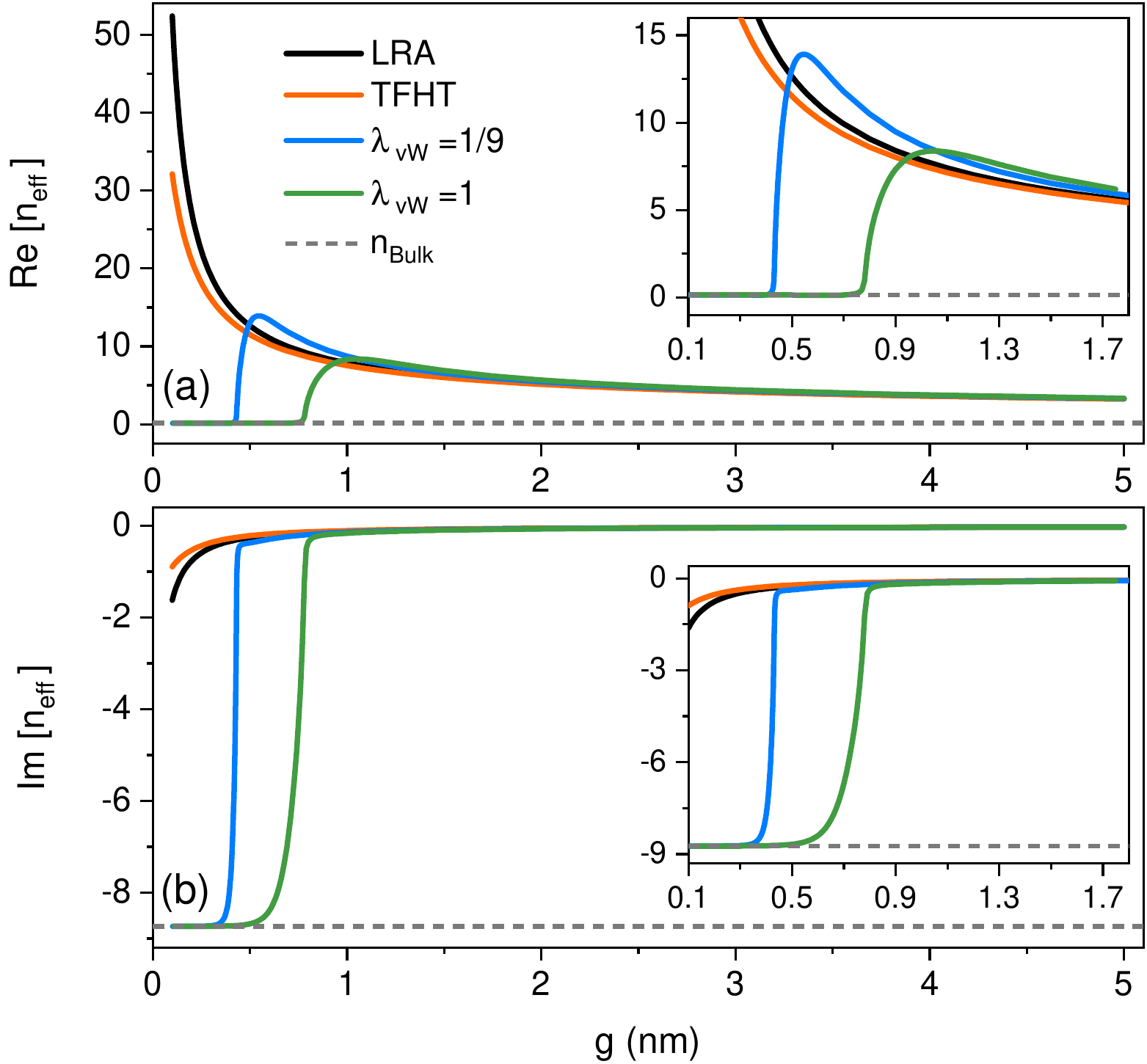}
	\caption{Effective refractive index for Ag-air-Ag configuration as a function of gap distance. A comparison between the LRA, TFHT and QHT is given. The upper and lower panels show the real and imaginary parts of the effective index, respectively. The results are plotted at $E=1$~eV. The dashed grey line represents the refractive index of the bulk metal.}
	\label{fig:Ag_gapSweep}
\end{figure}

Noble metals, such as Ag, show in general a more complex behavior compared to Drude-type metals due the contribution of interband transitions and core electrons.
In what follows, we use $r_s=3$~a.u. and $\hbar\gamma=0.03$~eV for Ag, and account for interband effects with a constant background dielectric constant $\varepsilon_\infty=5.9$ \cite{Khalid2020commphys}, which adds a local contribution to the free-electron metal permittivity. 
The effective refractive index for the Ag-air-Ag configuration as a function of the gap size computed using different models are plotted in Fig.~\ref{fig:Ag_gapSweep}.
The real and imaginary parts of $n_{\rm eff}$ are plotted in Fig.~\ref{fig:Ag_gapSweep}~(a) and (b), respectively, showing quite similar behavior as seen for the Na-Air-Na case in the previous section.
LRA and TFHT clearly show unphysical increasing character of effective index when sub-nanometer gaps are in question. QHT, on the other hand, predicts that in the tunneling regime, for sufficiently small metal separation, the gap virtually fades away completely and the effective index converges to the refractive index of the bulk Ag.
Note that in Fig.~\ref{fig:Ag_gapSweep}, the QHT gives a different threshold for $g$ at which the gap virtually vanishes, depending on the amount of electron spill-out considered from the metal surface, i.e., $\lambda_{\rm vW}=1/9$ or $\lambda_{\rm vW}=1$.    

\begin{figure}
	\centering
	\includegraphics[width=0.9\linewidth]{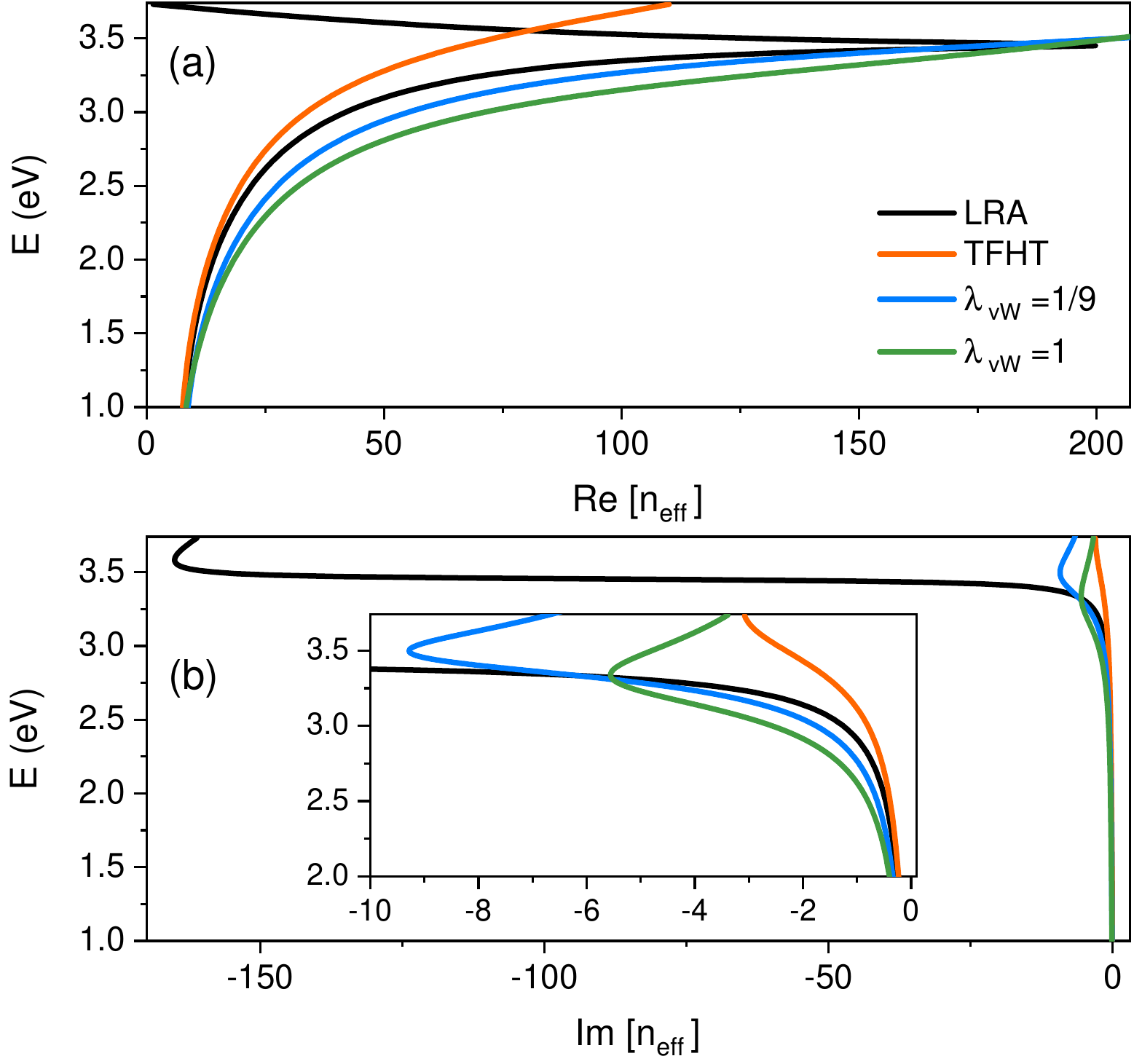}
	\caption{Effective refractive index as a function of energy (in eV) at $g=1~$nm. A comparison between the LRA and QHT is given. The upper and lower panels show the real and imaginary parts of the effective index. The results are normalized with the screened plasma frequency, $\omega_p'=\omega_p/\sqrt{\varepsilon_\infty}$.}
	\label{fig:Ag_Esweep}
\end{figure}

So far, we have analyzed the effective mode index of the MIM waveguide as function of air gap between the metals.
It is also important to examine the behavior of $n_{\rm eff}$ as a function of energy.
Figure~\ref{fig:Ag_Esweep} (a) and (b) present the real and imaginary parts of $n_{\rm eff}$, respectively, as a function of energy ($E$ in eV) for a Ag-air-Ag waveguide, considering a gap size $g=1$~nm -- a width which is experimentally rather easy to obtain \cite{Ciraci2012Science}.

This underlines the differences between the predictions of the different approaches regarding the effective index of the mode. The LRA predicts a bend back for the real part and a much larger imaginary part in this region. The TFHT on the contrary predicts a much lower imaginary part at any frequency. The QHT approach predicts a more reasonable behavior of the gap-plasmon with no bend back in the effective index and values of its imaginary part that can be much lower than for the LRA at higher frequencies. 

The lower the imaginary part of the effective index, the larger the typical propagation length of the gap-plasmon, and finally the larger the quality factor of an eventual cavity for the gap-plasmon \cite{lemaitre2017interferometric}. This underlines the importance of an accurate prediction in this regime. 

\section{Effective gap}\label{sec:effective}

To summarize qualitatively the above results, the spill-out seems to be able to close the gap electromagnetically speaking before it is actually spatially or mechanically closed. In addition, we have seen that when the spill-out is taken into account, whether in the simplified model or within the QHT framework, the effective index of the gap-plasmon is slightly larger than when the spill-out is neglected, however it presents a very similar trend. All of these results suggest that everything occurs as if the gap were narrower when the spill-out is taken into account, which seems physically sound. The fact that, for the simplified model, peaks seem to appear when the electron density reaches a certain value suggests that it should be possible to define the electromagnetic edges of the metal, as well as an effective value for the gap width. 
We define this value as the width that should be considered in the LRA model or in the TFHT approach for the gap-plasmon to fit the effective index of the gap-plasmon provided by the QHT. 

The effective gap is thus given by 
$$g_e = g - \delta_g$$
where $g$ is the actual gap between the edges of the metal. Said otherwise, in order to retrieve the results of the QHT, one can use the LRA (or the TFHT) with hard wall boundaries, but simply use a diminished size of the gap to take the spill-out into account. 

\begin{figure}
	\centering
	\includegraphics[width=0.9\linewidth]{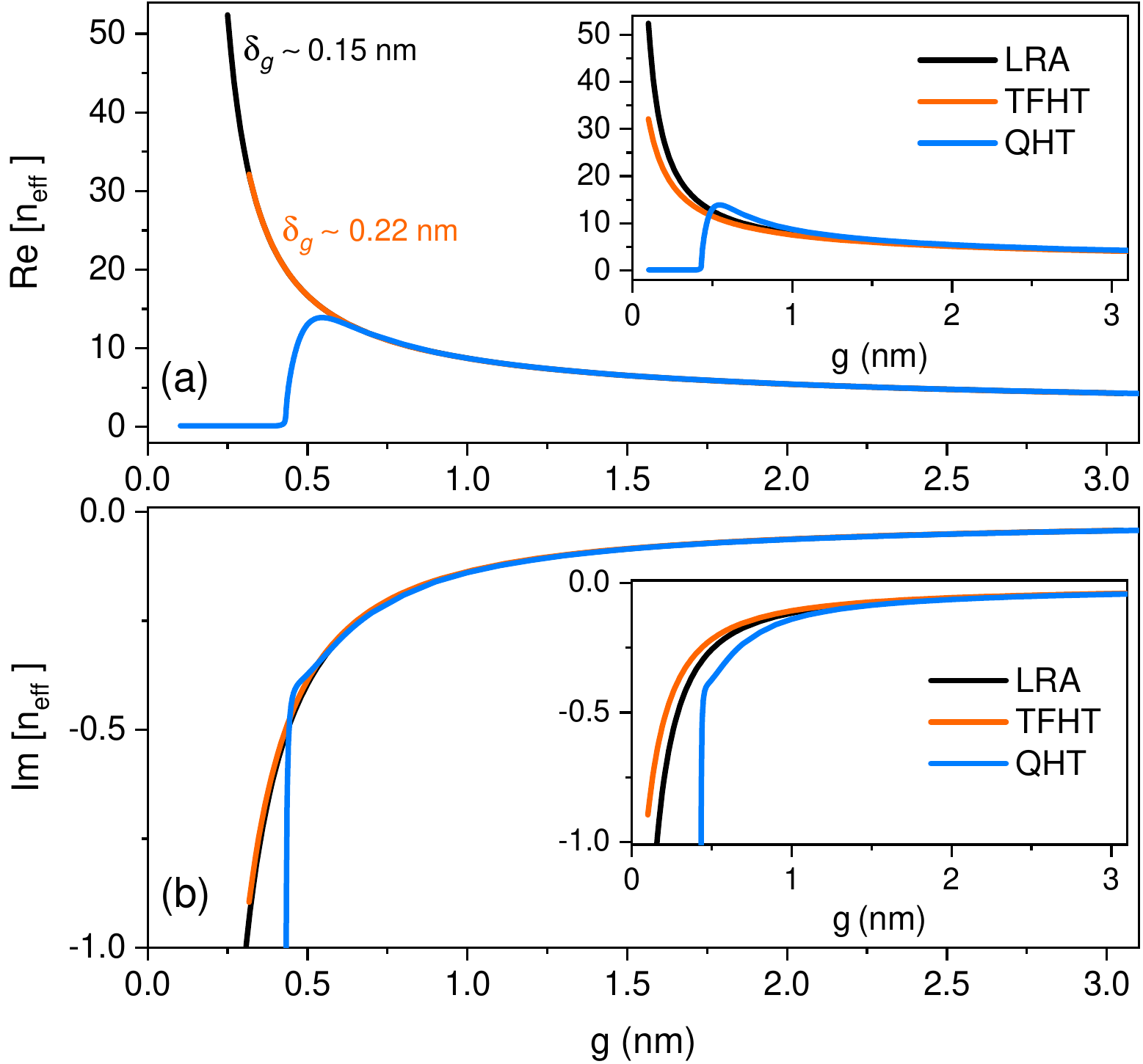}
	\caption{The effective mode index for Ag-air-Ag waveguide (a) real and (b) imaginary part as a function of gap distance $g$ between the metal edges plotted at $E=1~$eV. The LRA and TFHT curves are shifted by $\delta_g$ to cope with the QHT results, which allows to introduce the concept of an effective gap that can be defined as $g_e=g-\delta_g$. The $g_e$ seems to appear smaller than the mechanical gap $g$ by an amount of $\delta_g$. In particular, using $\delta_g=0.15$~nm in the LRA and $\delta_g=0.22$~nm in the TF approximation gives good fit with QHT. The insets plot $n_{\rm{eff}}$ as a function of actual gap.}
	\label{fig:effectiveGap}
\end{figure}

For silver, shifting the LRA curve of $\delta_g=0.15$ nm or the TFHT curve of $\delta_g=0.22$ nm allows to retrieve the effective index of the gap-plasmon as computed with the QHT extremely accurately, as long as the gap does not start closing (which occurs around $g \simeq 0.6$ nm). The results for the QHT reported in Fig.~\ref{fig:effectiveGap} are for $\lambda_{\rm vW}=1/9$, 
however, the QHT result for $\lambda_{\rm vW}=1$ can also be nicely reproduced from the LRA and TF approximations considering $\delta_g\simeq 0.25$~nm and $\delta_g\simeq 0.32$~nm, respectively. 
The idea that the gap-plasmon is sensitive to the effective gap between the metals is thus well founded, as it yields quantitative results. This can be understood because the spill out makes the response of the electron gas outside the metal similar to the response inside the metal, as soon as the electron density is high enough.
However, we point out that we have not found another way to define the effective gap based on the profile of the electron density or of the fields. The values we give are simply the ones that yield the best fit and there does not seem to be any simple way to predict such values.

This discussion finally underlines that at such scales, for the gap-plasmon, the tiny difference (a rule of thumb would be, according to our results, to consider a 0.1 nm difference) that exists between the actual edge of the metal and the ``electromagnetic edge" has its importance and must be considered in the limit of vanishing gaps. 

\section{Conclusions}\label{sec:conclusions}
In this article, we analyzed the impact of both electron spill-out and nonlocality on the propagation of gap-plasmons in MIM waveguides using  the full self-consistent QHT, which can efficiently take into account both spatial dispersion inside the electron gas and the spatial variation of the electron density outside of the metal. We also compare our results with the commonly used approaches (LRA and TFHT) without any spill-out and with a popular LRA approach for the spill out \cite{Esteban2012NatCommun,SkjolstrupPhys.Rev.B.,Skjolstrup2019_PRB}.

Peaks in the electric field profile and a large imaginary part for the effective index of the gap plasmon are predicted by the latter approach. Such features are however absent in the framework of the QHT, showing that nonlocality is key to an accurate description of the gap plasmon.
We underline that only in the QHT framework does the effective mode index converge to the refractive index of the bulk metal, as expected from physical considerations. 
Hence QHT, without being affected by the artifacts present in the local description, provides a good measure of the effective mode index and of the local field amplitudes, which are particularly relevant for nonlinear optical processes where response of the system is strongly affected by the local intensities.
These points reinforce the idea that QHT is the right tool for plasmonics at the subnanometer level while using a LRA model for the spill-out may lead, at least in the case of gap-plasmon resonators, to an overestimation of the losses and an underestimation of the quality factor of the resonance entirely due to the fact that nonlocality has been neglected. 

Our study thus shows that the spill-out indeed imposes an upper limit on the effective index of the gap-plasmon, confirming previous studies \cite{SkjolstrupPhys.Rev.B.}. A gap-plasmon is very unlikely to present an effective index larger than 15 for noble metals in the visible, which puts a theoretical limit to the miniaturization of gap-plasmon resonators in the visible.

Importantly, whatever model is chosen to describe the spill-out, it seems possible to introduce an effective gap width between metals, smaller than the gap size defined as the distance between the atoms of the metal across the gap. This effective gap width allows to take simply the effect of the spill-out into account, as long as the gap can be considered open to the propagation of the mode. In many experiments, the definition of the gap between metals is of crucial importance \cite{Ciraci2012Science,moreau2012controlled}. Our reasoning suggests using optical characterization methods (like ellipsometry typically) to determine the thickness of a material deposited at the surface of a metal is particularly relevant in the context of gap-plasmon resonators, as the optical thickness which is retrieved probably takes the spill-out partially into account. The gap width deduced from such measurements is in that way probably closer to the effective gap width we have defined here. This may thus explain why hard-wall approaches have been so far surprisingly successful at predicting the experimental behavior of gap-plasmons in the limit of vanishing gaps.

\begin{acknowledgments}
Antoine Moreau is an Academy CAP 20-25 chair holder. He acknowledges the support received from the Agence Nationale de la Recherche of the French government through the program “Investissements d’Avenir” (16-IDEX-0001 CAP 20-25).
\end{acknowledgments}

\appendix
\renewcommand\thefigure{\thesection.\arabic{figure}}  
\counterwithin{figure}{section}     
\section{Numerical Details}\label{sec:appendix}
\begin{figure}
\centering
\includegraphics[width=0.9\linewidth]{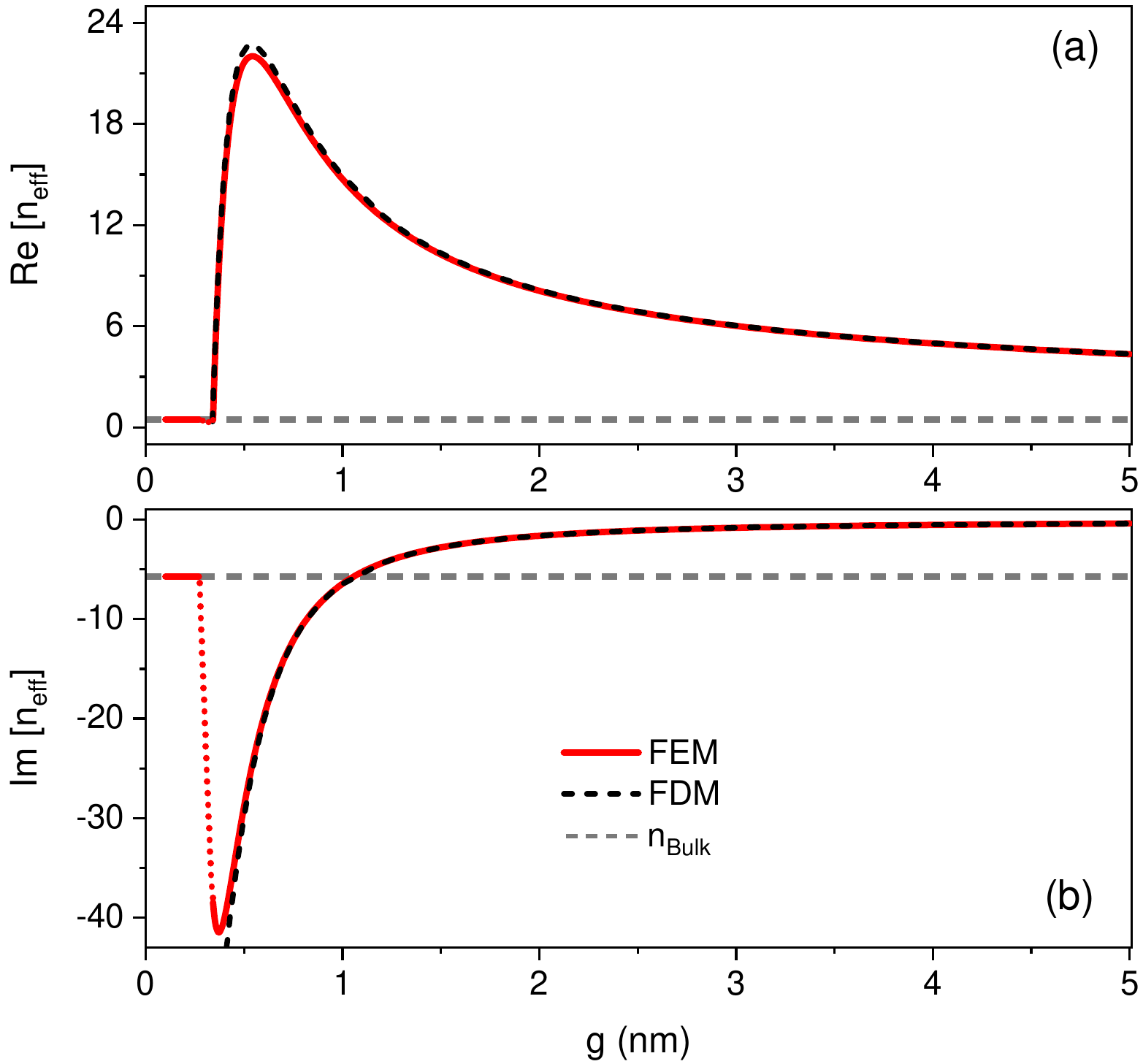}
\caption{Comparison between the finite-element-method (FEM) and finite-difference-method (FDM). (a) real and (b) imaginary parts of the effective refractive index $n_{\rm eff}$. The horizontal dashed-gery line represents the refractive index of the bulk Na.} \label{fig:FEM_FDM}
\end{figure}

We implement Eqs.~(\ref{eq:wave}), (\ref{eq:scn0}) and (\ref{eq:J}) in a finite-element-method based commercial simulator \textsc{comsol} Multiphysics. The complete set of equations in their weak formulation, recalling $\partial_t \mathbf{P}_f=\mathbf{J}$, can be expressed as:
\begin{widetext}
\begin{eqnarray}
& &\int \left(\nabla \times \boldsymbol{\mathrm{E}} \right) \cdot \left(\nabla \times \tilde{\boldsymbol{\mathrm{E}}} \right)-\left( \varepsilon_\infty \frac{\omega^2}{c^2}\boldsymbol{\mathrm{E}}+\mu_0 \omega^2 \boldsymbol{\mathrm{P}}_f \right)\cdot \tilde{\boldsymbol{\mathrm{E}}} ~dV=0, \label{eq:wWave}\\
& &\int -\frac{e}{m_e} \left(\frac{\delta G[n]}{\delta n} \right) \left(\nabla \cdot \tilde{\boldsymbol{\mathrm{P}}}_f\right)+\frac{1}{n_0}\left[\left(\omega^2+i \gamma\omega \right) \boldsymbol{\mathrm{P}}_f+\varepsilon_0 \omega_p^2\boldsymbol{\mathrm{E}}\right]\cdot \tilde{\boldsymbol{\mathrm{P}}}_f~dV=0, \label{eq:wP}\\
& &\int -\left(\nabla \cdot \boldsymbol{\mathrm{P}}_f \right)\left(\nabla \cdot \tilde{\boldsymbol{\mathrm{F}}} \right)-e \boldsymbol{\mathrm{F}} \cdot \tilde{\boldsymbol{\mathrm{F}}} ~dV=0, \label{eq:wF}\\
& &\int - \varepsilon_\infty \nabla \left(\frac{\delta G[\xi^2]}{\delta n} \right)_{\xi=\sqrt{n_0}} \cdot \nabla \tilde{\xi}+\frac{e^2}{\varepsilon_0}\left(\xi^2-n^+ \right)\tilde{\xi} ~dV=0, \label{eq:wn0}
\end{eqnarray}
\end{widetext}
where the field quantities with a tilde $(\sim)$ sign indicate the test functions. Since the energy functional contains the second-order derivatives, therefore, we introduce in Eq.~(\ref{eq:wF}) a working variable $\mathbf{F}$, such that, $\mathbf{F}=\nabla n_1$ where $n_1=\frac{1}{e}\nabla \cdot \mathbf{P}_f$. We compute the self-consistent equilibrium charge density $n_0$ from Eq.~(\ref{eq:wn0}) and we find that the solution converges more quickly by using a transformed variable $\xi=\sqrt{n_0}$. Equation~(\ref{eq:wP}) predicts the linear response of the system within the framework of QHT, with $G[n]$ as given by Eq.~(\ref{eq:G[n]}). Assuming $n_0(\mathbf{r})=n_0$ and $G[n]=T_{\rm TF}[n]$ with the hard-wall boundary condition gives the standard TFHT, and considering $G[n]=0$ yields the LRA.
In the simplified spill-out model given by Eq. \eqref{symp_mod}, we use the model density as given by Eq.~(\ref{eq:an_n0}), taking electron spill-out into account.
We have solved Eq.~\eqref{symp_mod} by an iterative procedure based on a finite-difference scheme. 
For thin gap, the decay length of the solution becomes much larger than the gap size. The presence of these two scales is the source of numerical instabilities. In order to obtain a accurate resolution of the length scale associated to the gap, we reduce the computation domain. This is obtained by introducing a artificial interface inside the metal at a distance $L$ from the origin. We solve the problem in the interval $[0,L]$ and we use the analytical solution at the interface as a boundary condition in the finite-difference scheme. In order to illustrate the numerical scheme, we focus on the term 
$  \partial_y \left(v(y) \partial_y E_z 
	\right)
$, where we have defined $v\doteq  
	\frac{k_z^2}{\varepsilon(y)\, k_0^2-k_z^2}$. 
The other terms are treated similarly. 
We discretize the interval $[0,L]$ by a equispaced grid with size $\Delta$. We apply a second order central scheme 
\begin{align*}
	\partial_y \left(v \partial_y E_z 
	\right)	(y_i)
	=&\frac{E_{i-1}v^-_i-E_{i}(v^+_i+v^-_i)+E_{i+1}v^+_i}{2\Delta^2}  , 
\end{align*}
where $E_i=E_z(y_i)$ 
 represents the field at the discrete position $y_i=(i-1)\Delta$ and $	v^\pm_i \doteq  v(y_{i\pm1}) +v(y_i) $. At the boundary,
\begin{align*}
 \partial_y \left(v \partial_y E_z 
\right) (L)
	=&\frac{v (L) \left(E_z(L-\Delta) +E_z(L)\left(  e^{ik_z \Delta }-2 \right)\right) }{\Delta^2}   ,
\end{align*}
where we use that for $y>L$ the field has the exponential form $E_z=E_z(L)e^{-ik_z (y-L)}$. 
We obtain a homogeneous linear system for the discrete field $E(y_i)$ parameterized by the wave number $k_z$. The plasmon resonances are associated to the $k_z $ for which the system has a non trivial kernel. We use a iterative procedure based on the Nelder-Mead simplex algorithm \cite{Lagarias1998} where we vary  $k_z$ in the complex plane in order to minimize the singular values of the matrix associated to the linear system.
The results are depicted in Fig. \ref{fig:FEM_FDM} where we show the agreement of the results obtained by the simplified model based on finite-difference scheme with the solution of Eqs. \eqref{eq:wWave}-\eqref{eq:wn0} obtained by a  finite-element-method.

\bibliography{main}

\begin{thebibliography}{47}%
\makeatletter
\providecommand \@ifxundefined [1]{%
 \@ifx{#1\undefined}
}%
\providecommand \@ifnum [1]{%
 \ifnum #1\expandafter \@firstoftwo
 \else \expandafter \@secondoftwo
 \fi
}%
\providecommand \@ifx [1]{%
 \ifx #1\expandafter \@firstoftwo
 \else \expandafter \@secondoftwo
 \fi
}%
\providecommand \natexlab [1]{#1}%
\providecommand \enquote  [1]{``#1''}%
\providecommand \bibnamefont  [1]{#1}%
\providecommand \bibfnamefont [1]{#1}%
\providecommand \citenamefont [1]{#1}%
\providecommand \href@noop [0]{\@secondoftwo}%
\providecommand \href [0]{\begingroup \@sanitize@url \@href}%
\providecommand \@href[1]{\@@startlink{#1}\@@href}%
\providecommand \@@href[1]{\endgroup#1\@@endlink}%
\providecommand \@sanitize@url [0]{\catcode `\\12\catcode `\$12\catcode
  `\&12\catcode `\#12\catcode `\^12\catcode `\_12\catcode `\%12\relax}%
\providecommand \@@startlink[1]{}%
\providecommand \@@endlink[0]{}%
\providecommand \url  [0]{\begingroup\@sanitize@url \@url }%
\providecommand \@url [1]{\endgroup\@href {#1}{\urlprefix }}%
\providecommand \urlprefix  [0]{URL }%
\providecommand \Eprint [0]{\href }%
\providecommand \doibase [0]{https://doi.org/}%
\providecommand \selectlanguage [0]{\@gobble}%
\providecommand \bibinfo  [0]{\@secondoftwo}%
\providecommand \bibfield  [0]{\@secondoftwo}%
\providecommand \translation [1]{[#1]}%
\providecommand \BibitemOpen [0]{}%
\providecommand \bibitemStop [0]{}%
\providecommand \bibitemNoStop [0]{.\EOS\space}%
\providecommand \EOS [0]{\spacefactor3000\relax}%
\providecommand \BibitemShut  [1]{\csname bibitem#1\endcsname}%
\let\auto@bib@innerbib\@empty
\bibitem [{\citenamefont {Baumberg}\ \emph {et~al.}(2019)\citenamefont
  {Baumberg}, \citenamefont {Aizpurua}, \citenamefont {Mikkelsen},\ and\
  \citenamefont {Smith}}]{baumberg2019extreme}%
  \BibitemOpen
  \bibfield  {author} {\bibinfo {author} {\bibfnamefont {J.~J.}\ \bibnamefont
  {Baumberg}}, \bibinfo {author} {\bibfnamefont {J.}~\bibnamefont {Aizpurua}},
  \bibinfo {author} {\bibfnamefont {M.~H.}\ \bibnamefont {Mikkelsen}},\ and\
  \bibinfo {author} {\bibfnamefont {D.~R.}\ \bibnamefont {Smith}},\ }\bibfield
  {title} {\bibinfo {title} {Extreme nanophotonics from ultrathin metallic
  gaps},\ }\href@noop {} {\bibfield  {journal} {\bibinfo  {journal} {Nature
  materials}\ }\textbf {\bibinfo {volume} {18}},\ \bibinfo {pages} {668}
  (\bibinfo {year} {2019})}\BibitemShut {NoStop}%
\bibitem [{\citenamefont {Wang}\ and\ \citenamefont
  {Kempa}(2007)}]{Wang2007Phys.Rev.B.}%
  \BibitemOpen
  \bibfield  {author} {\bibinfo {author} {\bibfnamefont {X.}~\bibnamefont
  {Wang}}\ and\ \bibinfo {author} {\bibfnamefont {K.}~\bibnamefont {Kempa}},\
  }\bibfield  {title} {\bibinfo {title} {Plasmon polaritons in slot waveguides:
  Simple model calculations and a full nonlocal quantum mechanical treatment},\
  }\href {https://doi.org/10.1103/PhysRevB.75.245426} {\bibfield  {journal}
  {\bibinfo  {journal} {Phys. Rev. B}\ }\textbf {\bibinfo {volume} {75}},\
  \bibinfo {pages} {245426} (\bibinfo {year} {2007})}\BibitemShut {NoStop}%
\bibitem [{\citenamefont {Jung}\ \emph {et~al.}(2009)\citenamefont {Jung},
  \citenamefont {S\o{}ndergaard},\ and\ \citenamefont
  {Bozhevolnyi}}]{Jung2009Phys.Rev.B.}%
  \BibitemOpen
  \bibfield  {author} {\bibinfo {author} {\bibfnamefont {J.}~\bibnamefont
  {Jung}}, \bibinfo {author} {\bibfnamefont {T.}~\bibnamefont
  {S\o{}ndergaard}},\ and\ \bibinfo {author} {\bibfnamefont {S.~I.}\
  \bibnamefont {Bozhevolnyi}},\ }\bibfield  {title} {\bibinfo {title} {Gap
  plasmon-polariton nanoresonators: Scattering enhancement and launching of
  surface plasmon polaritons},\ }\href
  {https://doi.org/10.1103/PhysRevB.79.035401} {\bibfield  {journal} {\bibinfo
  {journal} {Phys. Rev. B}\ }\textbf {\bibinfo {volume} {79}},\ \bibinfo
  {pages} {035401} (\bibinfo {year} {2009})}\BibitemShut {NoStop}%
\bibitem [{\citenamefont {Akselrod}\ \emph {et~al.}(2014)\citenamefont
  {Akselrod}, \citenamefont {Argyropoulos}, \citenamefont {Hoang},
  \citenamefont {Cirac{\`\i}}, \citenamefont {Fang}, \citenamefont {Huang},
  \citenamefont {Smith},\ and\ \citenamefont
  {Mikkelsen}}]{akselrod2014probing}%
  \BibitemOpen
  \bibfield  {author} {\bibinfo {author} {\bibfnamefont {G.~M.}\ \bibnamefont
  {Akselrod}}, \bibinfo {author} {\bibfnamefont {C.}~\bibnamefont
  {Argyropoulos}}, \bibinfo {author} {\bibfnamefont {T.~B.}\ \bibnamefont
  {Hoang}}, \bibinfo {author} {\bibfnamefont {C.}~\bibnamefont {Cirac{\`\i}}},
  \bibinfo {author} {\bibfnamefont {C.}~\bibnamefont {Fang}}, \bibinfo {author}
  {\bibfnamefont {J.}~\bibnamefont {Huang}}, \bibinfo {author} {\bibfnamefont
  {D.~R.}\ \bibnamefont {Smith}},\ and\ \bibinfo {author} {\bibfnamefont
  {M.~H.}\ \bibnamefont {Mikkelsen}},\ }\bibfield  {title} {\bibinfo {title}
  {Probing the mechanisms of large purcell enhancement in plasmonic
  nanoantennas},\ }\href@noop {} {\bibfield  {journal} {\bibinfo  {journal}
  {Nature Photonics}\ }\textbf {\bibinfo {volume} {8}},\ \bibinfo {pages} {835}
  (\bibinfo {year} {2014})}\BibitemShut {NoStop}%
\bibitem [{\citenamefont {Hoang}\ \emph {et~al.}(2015)\citenamefont {Hoang},
  \citenamefont {Akselrod}, \citenamefont {Argyropoulos}, \citenamefont
  {Huang}, \citenamefont {Smith},\ and\ \citenamefont
  {Mikkelsen}}]{hoang2015ultrafast}%
  \BibitemOpen
  \bibfield  {author} {\bibinfo {author} {\bibfnamefont {T.~B.}\ \bibnamefont
  {Hoang}}, \bibinfo {author} {\bibfnamefont {G.~M.}\ \bibnamefont {Akselrod}},
  \bibinfo {author} {\bibfnamefont {C.}~\bibnamefont {Argyropoulos}}, \bibinfo
  {author} {\bibfnamefont {J.}~\bibnamefont {Huang}}, \bibinfo {author}
  {\bibfnamefont {D.~R.}\ \bibnamefont {Smith}},\ and\ \bibinfo {author}
  {\bibfnamefont {M.~H.}\ \bibnamefont {Mikkelsen}},\ }\bibfield  {title}
  {\bibinfo {title} {Ultrafast spontaneous emission source using plasmonic
  nanoantennas},\ }\href@noop {} {\bibfield  {journal} {\bibinfo  {journal}
  {Nature communications}\ }\textbf {\bibinfo {volume} {6}},\ \bibinfo {pages}
  {1} (\bibinfo {year} {2015})}\BibitemShut {NoStop}%
\bibitem [{\citenamefont {Yang}\ \emph {et~al.}(2012)\citenamefont {Yang},
  \citenamefont {Sauvan}, \citenamefont {Jouanin}, \citenamefont {Collin},
  \citenamefont {Pelouard},\ and\ \citenamefont
  {Lalanne}}]{yang2012ultrasmall}%
  \BibitemOpen
  \bibfield  {author} {\bibinfo {author} {\bibfnamefont {J.}~\bibnamefont
  {Yang}}, \bibinfo {author} {\bibfnamefont {C.}~\bibnamefont {Sauvan}},
  \bibinfo {author} {\bibfnamefont {A.}~\bibnamefont {Jouanin}}, \bibinfo
  {author} {\bibfnamefont {S.}~\bibnamefont {Collin}}, \bibinfo {author}
  {\bibfnamefont {J.-L.}\ \bibnamefont {Pelouard}},\ and\ \bibinfo {author}
  {\bibfnamefont {P.}~\bibnamefont {Lalanne}},\ }\bibfield  {title} {\bibinfo
  {title} {Ultrasmall metal-insulator-metal nanoresonators: impact of slow-wave
  effects on the quality factor},\ }\href@noop {} {\bibfield  {journal}
  {\bibinfo  {journal} {Optics Express}\ }\textbf {\bibinfo {volume} {20}},\
  \bibinfo {pages} {16880} (\bibinfo {year} {2012})}\BibitemShut {NoStop}%
\bibitem [{\citenamefont {Moreau}\ \emph {et~al.}(2012)\citenamefont {Moreau},
  \citenamefont {Cirac{\`\i}}, \citenamefont {Mock}, \citenamefont {Hill},
  \citenamefont {Wang}, \citenamefont {Wiley}, \citenamefont {Chilkoti},\ and\
  \citenamefont {Smith}}]{moreau2012controlled}%
  \BibitemOpen
  \bibfield  {author} {\bibinfo {author} {\bibfnamefont {A.}~\bibnamefont
  {Moreau}}, \bibinfo {author} {\bibfnamefont {C.}~\bibnamefont {Cirac{\`\i}}},
  \bibinfo {author} {\bibfnamefont {J.~J.}\ \bibnamefont {Mock}}, \bibinfo
  {author} {\bibfnamefont {R.~T.}\ \bibnamefont {Hill}}, \bibinfo {author}
  {\bibfnamefont {Q.}~\bibnamefont {Wang}}, \bibinfo {author} {\bibfnamefont
  {B.~J.}\ \bibnamefont {Wiley}}, \bibinfo {author} {\bibfnamefont
  {A.}~\bibnamefont {Chilkoti}},\ and\ \bibinfo {author} {\bibfnamefont
  {D.~R.}\ \bibnamefont {Smith}},\ }\bibfield  {title} {\bibinfo {title}
  {Controlled-reflectance surfaces with film-coupled colloidal nanoantennas},\
  }\href@noop {} {\bibfield  {journal} {\bibinfo  {journal} {Nature}\ }\textbf
  {\bibinfo {volume} {492}},\ \bibinfo {pages} {86} (\bibinfo {year}
  {2012})}\BibitemShut {NoStop}%
\bibitem [{\citenamefont {S{\o}ndergaard}\ \emph {et~al.}(2012)\citenamefont
  {S{\o}ndergaard}, \citenamefont {Novikov}, \citenamefont {Holmgaard},
  \citenamefont {Eriksen}, \citenamefont {Beermann}, \citenamefont {Han},
  \citenamefont {Pedersen},\ and\ \citenamefont
  {Bozhevolnyi}}]{sondergaard2012plasmonic}%
  \BibitemOpen
  \bibfield  {author} {\bibinfo {author} {\bibfnamefont {T.}~\bibnamefont
  {S{\o}ndergaard}}, \bibinfo {author} {\bibfnamefont {S.~M.}\ \bibnamefont
  {Novikov}}, \bibinfo {author} {\bibfnamefont {T.}~\bibnamefont {Holmgaard}},
  \bibinfo {author} {\bibfnamefont {R.~L.}\ \bibnamefont {Eriksen}}, \bibinfo
  {author} {\bibfnamefont {J.}~\bibnamefont {Beermann}}, \bibinfo {author}
  {\bibfnamefont {Z.}~\bibnamefont {Han}}, \bibinfo {author} {\bibfnamefont
  {K.}~\bibnamefont {Pedersen}},\ and\ \bibinfo {author} {\bibfnamefont
  {S.~I.}\ \bibnamefont {Bozhevolnyi}},\ }\bibfield  {title} {\bibinfo {title}
  {Plasmonic black gold by adiabatic nanofocusing and absorption of light in
  ultra-sharp convex grooves},\ }\href@noop {} {\bibfield  {journal} {\bibinfo
  {journal} {Nature communications}\ }\textbf {\bibinfo {volume} {3}},\
  \bibinfo {pages} {1} (\bibinfo {year} {2012})}\BibitemShut {NoStop}%
\bibitem [{\citenamefont {Ciraci}\ \emph {et~al.}(2012)\citenamefont {Ciraci},
  \citenamefont {Hill}, \citenamefont {Mock}, \citenamefont {Urzhumov},
  \citenamefont {{Fernandez-Dominguez}}, \citenamefont {Maier}, \citenamefont
  {Pendry}, \citenamefont {Chilkoti},\ and\ \citenamefont
  {Smith}}]{Ciraci2012Science}%
  \BibitemOpen
  \bibfield  {author} {\bibinfo {author} {\bibfnamefont {C.}~\bibnamefont
  {Ciraci}}, \bibinfo {author} {\bibfnamefont {R.~T.}\ \bibnamefont {Hill}},
  \bibinfo {author} {\bibfnamefont {J.~J.}\ \bibnamefont {Mock}}, \bibinfo
  {author} {\bibfnamefont {Y.}~\bibnamefont {Urzhumov}}, \bibinfo {author}
  {\bibfnamefont {A.~I.}\ \bibnamefont {{Fernandez-Dominguez}}}, \bibinfo
  {author} {\bibfnamefont {S.~A.}\ \bibnamefont {Maier}}, \bibinfo {author}
  {\bibfnamefont {J.~B.}\ \bibnamefont {Pendry}}, \bibinfo {author}
  {\bibfnamefont {A.}~\bibnamefont {Chilkoti}},\ and\ \bibinfo {author}
  {\bibfnamefont {D.~R.}\ \bibnamefont {Smith}},\ }\bibfield  {title}
  {{\selectlanguage {english}\bibinfo {title} {Probing the {{Ultimate Limits}}
  of {{Plasmonic Enhancement}}}},\ }\href
  {https://doi.org/10.1126/science.1224823} {\bibfield  {journal} {\bibinfo
  {journal} {Science}\ }\textbf {\bibinfo {volume} {337}},\ \bibinfo {pages}
  {1072} (\bibinfo {year} {2012})}\BibitemShut {NoStop}%
\bibitem [{\citenamefont {Raza}\ \emph {et~al.}(2011)\citenamefont {Raza},
  \citenamefont {Toscano}, \citenamefont {Jauho}, \citenamefont {Wubs},\ and\
  \citenamefont {Mortensen}}]{Raza2011PhysRevB}%
  \BibitemOpen
  \bibfield  {author} {\bibinfo {author} {\bibfnamefont {S.}~\bibnamefont
  {Raza}}, \bibinfo {author} {\bibfnamefont {G.}~\bibnamefont {Toscano}},
  \bibinfo {author} {\bibfnamefont {A.-P.}\ \bibnamefont {Jauho}}, \bibinfo
  {author} {\bibfnamefont {M.}~\bibnamefont {Wubs}},\ and\ \bibinfo {author}
  {\bibfnamefont {N.~A.}\ \bibnamefont {Mortensen}},\ }\bibfield  {title}
  {\bibinfo {title} {Unusual resonances in nanoplasmonic structures due to
  nonlocal response},\ }\href {https://doi.org/10.1103/PhysRevB.84.121412}
  {\bibfield  {journal} {\bibinfo  {journal} {Phys. Rev. B}\ }\textbf {\bibinfo
  {volume} {84}},\ \bibinfo {pages} {121412} (\bibinfo {year}
  {2011})}\BibitemShut {NoStop}%
\bibitem [{\citenamefont {Cirac{\`i}}\ \emph {et~al.}(2013)\citenamefont
  {Cirac{\`i}}, \citenamefont {Pendry},\ and\ \citenamefont
  {Smith}}]{Ciraci2013ChemPhysChem}%
  \BibitemOpen
  \bibfield  {author} {\bibinfo {author} {\bibfnamefont {C.}~\bibnamefont
  {Cirac{\`i}}}, \bibinfo {author} {\bibfnamefont {J.~B.}\ \bibnamefont
  {Pendry}},\ and\ \bibinfo {author} {\bibfnamefont {D.~R.}\ \bibnamefont
  {Smith}},\ }\bibfield  {title} {{\selectlanguage {english}\bibinfo {title}
  {Hydrodynamic {{Model}} for {{Plasmonics}}: {{A Macroscopic Approach}} to a
  {{Microscopic Problem}}}},\ }\href {https://doi.org/10.1002/cphc.201200992}
  {\bibfield  {journal} {\bibinfo  {journal} {ChemPhysChem}\ }\textbf {\bibinfo
  {volume} {14}},\ \bibinfo {pages} {1109} (\bibinfo {year}
  {2013})}\BibitemShut {NoStop}%
\bibitem [{\citenamefont {Teperik}\ \emph {et~al.}(2013)\citenamefont
  {Teperik}, \citenamefont {Nordlander}, \citenamefont {Aizpurua},\ and\
  \citenamefont {Borisov}}]{Teperik2013Phys.Rev.Lett.}%
  \BibitemOpen
  \bibfield  {author} {\bibinfo {author} {\bibfnamefont {T.~V.}\ \bibnamefont
  {Teperik}}, \bibinfo {author} {\bibfnamefont {P.}~\bibnamefont {Nordlander}},
  \bibinfo {author} {\bibfnamefont {J.}~\bibnamefont {Aizpurua}},\ and\
  \bibinfo {author} {\bibfnamefont {A.~G.}\ \bibnamefont {Borisov}},\
  }\bibfield  {title} {{\selectlanguage {english}\bibinfo {title} {Robust
  {{Subnanometric Plasmon Ruler}} by {{Rescaling}} of the {{Nonlocal Optical
  Response}}}},\ }\href {https://doi.org/10.1103/PhysRevLett.110.263901}
  {\bibfield  {journal} {\bibinfo  {journal} {Phys. Rev. Lett.}\ }\textbf
  {\bibinfo {volume} {110}},\ \bibinfo {pages} {263901} (\bibinfo {year}
  {2013})}\BibitemShut {NoStop}%
\bibitem [{\citenamefont {Stella}\ \emph {et~al.}(2013)\citenamefont {Stella},
  \citenamefont {Zhang}, \citenamefont {García-Vidal}, \citenamefont {Rubio},\
  and\ \citenamefont {García-González}}]{Stella2013Phys.Chem.}%
  \BibitemOpen
  \bibfield  {author} {\bibinfo {author} {\bibfnamefont {L.}~\bibnamefont
  {Stella}}, \bibinfo {author} {\bibfnamefont {P.}~\bibnamefont {Zhang}},
  \bibinfo {author} {\bibfnamefont {F.~J.}\ \bibnamefont {García-Vidal}},
  \bibinfo {author} {\bibfnamefont {A.}~\bibnamefont {Rubio}},\ and\ \bibinfo
  {author} {\bibfnamefont {P.}~\bibnamefont {García-González}},\ }\bibfield
  {title} {\bibinfo {title} {Performance of nonlocal optics when applied to
  plasmonic nanostructures},\ }\href {https://doi.org/10.1021/jp401887y}
  {\bibfield  {journal} {\bibinfo  {journal} {The Journal of Physical Chemistry
  C}\ }\textbf {\bibinfo {volume} {117}},\ \bibinfo {pages} {8941} (\bibinfo
  {year} {2013})},\ \Eprint
  {https://arxiv.org/abs/https://doi.org/10.1021/jp401887y}
  {https://doi.org/10.1021/jp401887y} \BibitemShut {NoStop}%
\bibitem [{\citenamefont {Zhu}\ \emph {et~al.}(2016)\citenamefont {Zhu},
  \citenamefont {Esteban}, \citenamefont {Borisov}, \citenamefont {Baumberg},
  \citenamefont {Nordlander}, \citenamefont {Lezec}, \citenamefont {Aizpurua},\
  and\ \citenamefont {Crozier}}]{Zhu2016Nat.Com.}%
  \BibitemOpen
  \bibfield  {author} {\bibinfo {author} {\bibfnamefont {W.}~\bibnamefont
  {Zhu}}, \bibinfo {author} {\bibfnamefont {R.}~\bibnamefont {Esteban}},
  \bibinfo {author} {\bibfnamefont {A.~G.}\ \bibnamefont {Borisov}}, \bibinfo
  {author} {\bibfnamefont {J.~J.}\ \bibnamefont {Baumberg}}, \bibinfo {author}
  {\bibfnamefont {P.}~\bibnamefont {Nordlander}}, \bibinfo {author}
  {\bibfnamefont {H.~J.}\ \bibnamefont {Lezec}}, \bibinfo {author}
  {\bibfnamefont {J.}~\bibnamefont {Aizpurua}},\ and\ \bibinfo {author}
  {\bibfnamefont {K.~B.}\ \bibnamefont {Crozier}},\ }\bibfield  {title}
  {\bibinfo {title} {Quantum mechanical effects in plasmonic structures with
  subnanometre gaps},\ }\href {https://doi.org/10.1038/ncomms11495} {\bibfield
  {journal} {\bibinfo  {journal} {Nature Communications}\ }\textbf {\bibinfo
  {volume} {7}},\ \bibinfo {pages} {11495} (\bibinfo {year}
  {2016})}\BibitemShut {NoStop}%
\bibitem [{\citenamefont {Zuloaga}\ \emph {et~al.}(2009)\citenamefont
  {Zuloaga}, \citenamefont {Prodan},\ and\ \citenamefont
  {Nordlander}}]{Zuloaga2009}%
  \BibitemOpen
  \bibfield  {author} {\bibinfo {author} {\bibfnamefont {J.}~\bibnamefont
  {Zuloaga}}, \bibinfo {author} {\bibfnamefont {E.}~\bibnamefont {Prodan}},\
  and\ \bibinfo {author} {\bibfnamefont {P.}~\bibnamefont {Nordlander}},\
  }\bibfield  {title} {\bibinfo {title} {Quantum description of the plasmon
  resonances of a nanoparticle dimer},\ }\href
  {https://doi.org/10.1021/nl803811g} {\bibfield  {journal} {\bibinfo
  {journal} {Nano Letters}\ }\textbf {\bibinfo {volume} {9}},\ \bibinfo {pages}
  {887} (\bibinfo {year} {2009})},\ \bibinfo {note} {pMID: 19159319},\ \Eprint
  {https://arxiv.org/abs/https://doi.org/10.1021/nl803811g}
  {https://doi.org/10.1021/nl803811g} \BibitemShut {NoStop}%
\bibitem [{\citenamefont {Ullrich}(2011)}]{Ullrich2011}%
  \BibitemOpen
  \bibfield  {author} {\bibinfo {author} {\bibfnamefont {C.~A.}\ \bibnamefont
  {Ullrich}},\ }\href@noop {} {\emph {\bibinfo {title} {Time-Dependent
  Density-Functional Theory: Concepts and Applications}}}\ (\bibinfo
  {publisher} {Oxford University Press},\ \bibinfo {address} {Oxford},\
  \bibinfo {year} {2011})\BibitemShut {NoStop}%
\bibitem [{\citenamefont {Aguirregabiria}\ \emph {et~al.}(2018)\citenamefont
  {Aguirregabiria}, \citenamefont {Marinica}, \citenamefont {Esteban},
  \citenamefont {Kazansky}, \citenamefont {Aizpurua},\ and\ \citenamefont
  {Borisov}}]{Aguirregabiria2018}%
  \BibitemOpen
  \bibfield  {author} {\bibinfo {author} {\bibfnamefont {G.}~\bibnamefont
  {Aguirregabiria}}, \bibinfo {author} {\bibfnamefont {D.~C.}\ \bibnamefont
  {Marinica}}, \bibinfo {author} {\bibfnamefont {R.}~\bibnamefont {Esteban}},
  \bibinfo {author} {\bibfnamefont {A.~K.}\ \bibnamefont {Kazansky}}, \bibinfo
  {author} {\bibfnamefont {J.}~\bibnamefont {Aizpurua}},\ and\ \bibinfo
  {author} {\bibfnamefont {A.~G.}\ \bibnamefont {Borisov}},\ }\bibfield
  {title} {\bibinfo {title} {Role of electron tunneling in the nonlinear
  response of plasmonic nanogaps},\ }\href
  {https://doi.org/10.1103/PhysRevB.97.115430} {\bibfield  {journal} {\bibinfo
  {journal} {Phys. Rev. B}\ }\textbf {\bibinfo {volume} {97}},\ \bibinfo
  {pages} {115430} (\bibinfo {year} {2018})}\BibitemShut {NoStop}%
\bibitem [{\citenamefont {Esteban}\ \emph {et~al.}(2012)\citenamefont
  {Esteban}, \citenamefont {Borisov}, \citenamefont {Nordlander},\ and\
  \citenamefont {Aizpurua}}]{Esteban2012NatCommun}%
  \BibitemOpen
  \bibfield  {author} {\bibinfo {author} {\bibfnamefont {R.}~\bibnamefont
  {Esteban}}, \bibinfo {author} {\bibfnamefont {A.~G.}\ \bibnamefont
  {Borisov}}, \bibinfo {author} {\bibfnamefont {P.}~\bibnamefont
  {Nordlander}},\ and\ \bibinfo {author} {\bibfnamefont {J.}~\bibnamefont
  {Aizpurua}},\ }\bibfield  {title} {{\selectlanguage {english}\bibinfo {title}
  {Bridging quantum and classical plasmonics with a quantum-corrected model}},\
  }\href {https://doi.org/10.1038/ncomms1806} {\bibfield  {journal} {\bibinfo
  {journal} {Nat Commun}\ }\textbf {\bibinfo {volume} {3}},\ \bibinfo {pages}
  {825} (\bibinfo {year} {2012})}\BibitemShut {NoStop}%
\bibitem [{\citenamefont {Feiginov}\ and\ \citenamefont
  {Volkov}(1998)}]{Feiginov1998JETPLett}%
  \BibitemOpen
  \bibfield  {author} {\bibinfo {author} {\bibfnamefont {M.~N.}\ \bibnamefont
  {Feiginov}}\ and\ \bibinfo {author} {\bibfnamefont {V.~A.}\ \bibnamefont
  {Volkov}},\ }\bibfield  {title} {\bibinfo {title} {Self-excitation of {{2D}}
  plasmons in resonant tunneling diodes},\ }\href
  {https://doi.org/10.1134/1.567925} {\bibfield  {journal} {\bibinfo  {journal}
  {Journal of Experimental and Theoretical Physics Letters}\ }\textbf {\bibinfo
  {volume} {68}},\ \bibinfo {pages} {662} (\bibinfo {year} {1998})}\BibitemShut
  {NoStop}%
\bibitem [{\citenamefont {Ryzhii}\ and\ \citenamefont
  {Shur}(2001)}]{Ryzhii2001JJAppPhys}%
  \BibitemOpen
  \bibfield  {author} {\bibinfo {author} {\bibfnamefont {V.}~\bibnamefont
  {Ryzhii}}\ and\ \bibinfo {author} {\bibfnamefont {M.}~\bibnamefont {Shur}},\
  }\bibfield  {title} {\bibinfo {title} {Plasma instability and nonlinear
  terahertz oscillations in resonant-tunneling structures},\ }\href
  {https://doi.org/10.1143/jjap.40.546} {\bibfield  {journal} {\bibinfo
  {journal} {Japanese Journal of Applied Physics}\ }\textbf {\bibinfo {volume}
  {40}},\ \bibinfo {pages} {546} (\bibinfo {year} {2001})}\BibitemShut
  {NoStop}%
\bibitem [{\citenamefont {Svintsov}\ \emph {et~al.}(2016)\citenamefont
  {Svintsov}, \citenamefont {Devizorova}, \citenamefont {Otsuji},\ and\
  \citenamefont {Ryzhii}}]{Svintsov2016PhysRevB}%
  \BibitemOpen
  \bibfield  {author} {\bibinfo {author} {\bibfnamefont {D.}~\bibnamefont
  {Svintsov}}, \bibinfo {author} {\bibfnamefont {Z.}~\bibnamefont
  {Devizorova}}, \bibinfo {author} {\bibfnamefont {T.}~\bibnamefont {Otsuji}},\
  and\ \bibinfo {author} {\bibfnamefont {V.}~\bibnamefont {Ryzhii}},\
  }\bibfield  {title} {\bibinfo {title} {Plasmons in tunnel-coupled graphene
  layers: Backward waves with quantum cascade gain},\ }\href
  {https://doi.org/10.1103/PhysRevB.94.115301} {\bibfield  {journal} {\bibinfo
  {journal} {Phys. Rev. B}\ }\textbf {\bibinfo {volume} {94}},\ \bibinfo
  {pages} {115301} (\bibinfo {year} {2016})}\BibitemShut {NoStop}%
\bibitem [{\citenamefont {Skj\o{}lstrup}\ \emph {et~al.}(2018)\citenamefont
  {Skj\o{}lstrup}, \citenamefont {S\o{}ndergaard},\ and\ \citenamefont
  {Pedersen}}]{SkjolstrupPhys.Rev.B.}%
  \BibitemOpen
  \bibfield  {author} {\bibinfo {author} {\bibfnamefont {E.~J.~H.}\
  \bibnamefont {Skj\o{}lstrup}}, \bibinfo {author} {\bibfnamefont
  {T.}~\bibnamefont {S\o{}ndergaard}},\ and\ \bibinfo {author} {\bibfnamefont
  {T.~G.}\ \bibnamefont {Pedersen}},\ }\bibfield  {title} {\bibinfo {title}
  {Quantum spill-out in few-nanometer metal gaps: Effect on gap plasmons and
  reflectance from ultrasharp groove arrays},\ }\href
  {https://doi.org/10.1103/PhysRevB.97.115429} {\bibfield  {journal} {\bibinfo
  {journal} {Phys. Rev. B}\ }\textbf {\bibinfo {volume} {97}},\ \bibinfo
  {pages} {115429} (\bibinfo {year} {2018})}\BibitemShut {NoStop}%
\bibitem [{\citenamefont {Skj\o{}lstrup}\ \emph {et~al.}(2019)\citenamefont
  {Skj\o{}lstrup}, \citenamefont {S\o{}ndergaard},\ and\ \citenamefont
  {Pedersen}}]{Skjolstrup2019_PRB}%
  \BibitemOpen
  \bibfield  {author} {\bibinfo {author} {\bibfnamefont {E.~J.~H.}\
  \bibnamefont {Skj\o{}lstrup}}, \bibinfo {author} {\bibfnamefont
  {T.}~\bibnamefont {S\o{}ndergaard}},\ and\ \bibinfo {author} {\bibfnamefont
  {T.~G.}\ \bibnamefont {Pedersen}},\ }\bibfield  {title} {\bibinfo {title}
  {Quantum spill-out in nanometer-thin gold slabs: Effect on the plasmon mode
  index and the plasmonic absorption},\ }\href
  {https://doi.org/10.1103/PhysRevB.99.155427} {\bibfield  {journal} {\bibinfo
  {journal} {Phys. Rev. B}\ }\textbf {\bibinfo {volume} {99}},\ \bibinfo
  {pages} {155427} (\bibinfo {year} {2019})}\BibitemShut {NoStop}%
\bibitem [{\citenamefont {Toscano}\ \emph {et~al.}(2015)\citenamefont
  {Toscano}, \citenamefont {Straubel}, \citenamefont {Kwiatkowski},
  \citenamefont {Rockstuhl}, \citenamefont {Evers}, \citenamefont {Xu},
  \citenamefont {Asger~Mortensen},\ and\ \citenamefont
  {Wubs}}]{Toscano2015NatCommun}%
  \BibitemOpen
  \bibfield  {author} {\bibinfo {author} {\bibfnamefont {G.}~\bibnamefont
  {Toscano}}, \bibinfo {author} {\bibfnamefont {J.}~\bibnamefont {Straubel}},
  \bibinfo {author} {\bibfnamefont {A.}~\bibnamefont {Kwiatkowski}}, \bibinfo
  {author} {\bibfnamefont {C.}~\bibnamefont {Rockstuhl}}, \bibinfo {author}
  {\bibfnamefont {F.}~\bibnamefont {Evers}}, \bibinfo {author} {\bibfnamefont
  {H.}~\bibnamefont {Xu}}, \bibinfo {author} {\bibfnamefont {N.}~\bibnamefont
  {Asger~Mortensen}},\ and\ \bibinfo {author} {\bibfnamefont {M.}~\bibnamefont
  {Wubs}},\ }\bibfield  {title} {{\selectlanguage {english}\bibinfo {title}
  {Resonance shifts and spill-out effects in self-consistent hydrodynamic
  nanoplasmonics}},\ }\href {https://doi.org/10.1038/ncomms8132} {\bibfield
  {journal} {\bibinfo  {journal} {Nat Commun}\ }\textbf {\bibinfo {volume}
  {6}},\ \bibinfo {pages} {7132} (\bibinfo {year} {2015})}\BibitemShut
  {NoStop}%
\bibitem [{\citenamefont {Yan}(2015)}]{Yan2015Phys.Rev.B}%
  \BibitemOpen
  \bibfield  {author} {\bibinfo {author} {\bibfnamefont {W.}~\bibnamefont
  {Yan}},\ }\bibfield  {title} {{\selectlanguage {english}\bibinfo {title}
  {Hydrodynamic theory for quantum plasmonics: {{Linear}}-response dynamics of
  the inhomogeneous electron gas}},\ }\href
  {https://doi.org/10.1103/PhysRevB.91.115416} {\bibfield  {journal} {\bibinfo
  {journal} {Phys. Rev. B}\ }\textbf {\bibinfo {volume} {91}},\ \bibinfo
  {pages} {115416} (\bibinfo {year} {2015})}\BibitemShut {NoStop}%
\bibitem [{\citenamefont {Cirac{\`i}}\ and\ \citenamefont {{Della
  Sala}}(2016)}]{Ciraci2016Phys.Rev.B}%
  \BibitemOpen
  \bibfield  {author} {\bibinfo {author} {\bibfnamefont {C.}~\bibnamefont
  {Cirac{\`i}}}\ and\ \bibinfo {author} {\bibfnamefont {F.}~\bibnamefont
  {{Della Sala}}},\ }\bibfield  {title} {{\selectlanguage {english}\bibinfo
  {title} {Quantum hydrodynamic theory for plasmonics: {{Impact}} of the
  electron density tail}},\ }\href {https://doi.org/10.1103/PhysRevB.93.205405}
  {\bibfield  {journal} {\bibinfo  {journal} {Phys. Rev. B}\ }\textbf {\bibinfo
  {volume} {93}},\ \bibinfo {pages} {205405} (\bibinfo {year}
  {2016})}\BibitemShut {NoStop}%
\bibitem [{\citenamefont {Cirac{\`i}}(2017)}]{Ciraci2017Phys.Rev.B}%
  \BibitemOpen
  \bibfield  {author} {\bibinfo {author} {\bibfnamefont {C.}~\bibnamefont
  {Cirac{\`i}}},\ }\bibfield  {title} {{\selectlanguage {english}\bibinfo
  {title} {Current-dependent potential for nonlocal absorption in quantum
  hydrodynamic theory}},\ }\href {https://doi.org/10.1103/PhysRevB.95.245434}
  {\bibfield  {journal} {\bibinfo  {journal} {Phys. Rev. B}\ }\textbf {\bibinfo
  {volume} {95}},\ \bibinfo {pages} {245434} (\bibinfo {year}
  {2017})}\BibitemShut {NoStop}%
\bibitem [{\citenamefont {Ding}\ and\ \citenamefont
  {Chan}(2017)}]{Ding2017PRB}%
  \BibitemOpen
  \bibfield  {author} {\bibinfo {author} {\bibfnamefont {K.}~\bibnamefont
  {Ding}}\ and\ \bibinfo {author} {\bibfnamefont {C.~T.}\ \bibnamefont
  {Chan}},\ }\bibfield  {title} {\bibinfo {title} {Plasmonic modes of polygonal
  rods calculated using a quantum hydrodynamics method},\ }\href
  {https://doi.org/10.1103/PhysRevB.96.125134} {\bibfield  {journal} {\bibinfo
  {journal} {Phys. Rev. B}\ }\textbf {\bibinfo {volume} {96}},\ \bibinfo
  {pages} {125134} (\bibinfo {year} {2017})}\BibitemShut {NoStop}%
\bibitem [{\citenamefont {Gangaraj}\ and\ \citenamefont
  {Monticone}(2019)}]{Gangaraj2019Optica}%
  \BibitemOpen
  \bibfield  {author} {\bibinfo {author} {\bibfnamefont {S.~A.~H.}\
  \bibnamefont {Gangaraj}}\ and\ \bibinfo {author} {\bibfnamefont
  {F.}~\bibnamefont {Monticone}},\ }\bibfield  {title} {\bibinfo {title} {Do
  truly unidirectional surface plasmon-polaritons exist?},\ }\href
  {https://doi.org/10.1364/OPTICA.6.001158} {\bibfield  {journal} {\bibinfo
  {journal} {Optica}\ }\textbf {\bibinfo {volume} {6}},\ \bibinfo {pages}
  {1158} (\bibinfo {year} {2019})}\BibitemShut {NoStop}%
\bibitem [{\citenamefont {Hassani~Gangaraj}\ \emph {et~al.}(2020)\citenamefont
  {Hassani~Gangaraj}, \citenamefont {Jin}, \citenamefont {Argyropoulos},\ and\
  \citenamefont {Monticone}}]{Gangaraj2020PhysRevApp}%
  \BibitemOpen
  \bibfield  {author} {\bibinfo {author} {\bibfnamefont {S.~A.}\ \bibnamefont
  {Hassani~Gangaraj}}, \bibinfo {author} {\bibfnamefont {B.}~\bibnamefont
  {Jin}}, \bibinfo {author} {\bibfnamefont {C.}~\bibnamefont {Argyropoulos}},\
  and\ \bibinfo {author} {\bibfnamefont {F.}~\bibnamefont {Monticone}},\
  }\bibfield  {title} {\bibinfo {title} {Broadband field enhancement and giant
  nonlinear effects in terminated unidirectional plasmonic waveguides},\ }\href
  {https://doi.org/10.1103/PhysRevApplied.14.054061} {\bibfield  {journal}
  {\bibinfo  {journal} {Phys. Rev. Applied}\ }\textbf {\bibinfo {volume}
  {14}},\ \bibinfo {pages} {054061} (\bibinfo {year} {2020})}\BibitemShut
  {NoStop}%
\bibitem [{\citenamefont {Khalid}\ and\ \citenamefont
  {Ciracì}(2020)}]{Khalid2020commphys}%
  \BibitemOpen
  \bibfield  {author} {\bibinfo {author} {\bibfnamefont {M.}~\bibnamefont
  {Khalid}}\ and\ \bibinfo {author} {\bibfnamefont {C.}~\bibnamefont
  {Ciracì}},\ }\bibfield  {title} {\bibinfo {title} {Enhancing second-harmonic
  generation with electron spill-out at metallic surfaces},\ }\href
  {https://doi.org/https://doi.org/10.1038/s42005-020-00477-0} {\bibfield
  {journal} {\bibinfo  {journal} {Communications Physics}\ }\textbf {\bibinfo
  {volume} {3}},\ \bibinfo {pages} {214} (\bibinfo {year} {2020})}\BibitemShut
  {NoStop}%
\bibitem [{\citenamefont {Runge}\ and\ \citenamefont
  {Gross}(1984)}]{Runge.1984}%
  \BibitemOpen
  \bibfield  {author} {\bibinfo {author} {\bibfnamefont {E.}~\bibnamefont
  {Runge}}\ and\ \bibinfo {author} {\bibfnamefont {E.~K.~U.}\ \bibnamefont
  {Gross}},\ }\bibfield  {title} {\bibinfo {title} {{Density-Functional Theory
  for Time-Dependent Systems}},\ }\href
  {https://doi.org/10.1103/physrevlett.52.997} {\bibfield  {journal} {\bibinfo
  {journal} {Physical Review Letters}\ }\textbf {\bibinfo {volume} {52}},\
  \bibinfo {pages} {997} (\bibinfo {year} {1984})}\BibitemShut {NoStop}%
\bibitem [{\citenamefont {Manfredi}\ and\ \citenamefont
  {Haas}(2001)}]{manfredi2001self}%
  \BibitemOpen
  \bibfield  {author} {\bibinfo {author} {\bibfnamefont {G.}~\bibnamefont
  {Manfredi}}\ and\ \bibinfo {author} {\bibfnamefont {F.}~\bibnamefont
  {Haas}},\ }\bibfield  {title} {\bibinfo {title} {Self-consistent fluid model
  for a quantum electron gas},\ }\href@noop {} {\bibfield  {journal} {\bibinfo
  {journal} {Physical Review B}\ }\textbf {\bibinfo {volume} {64}},\ \bibinfo
  {pages} {075316} (\bibinfo {year} {2001})}\BibitemShut {NoStop}%
\bibitem [{\citenamefont {Crouseilles}\ \emph {et~al.}(2008)\citenamefont
  {Crouseilles}, \citenamefont {Hervieux},\ and\ \citenamefont
  {Manfredi}}]{crouseilles2008quantum}%
  \BibitemOpen
  \bibfield  {author} {\bibinfo {author} {\bibfnamefont {N.}~\bibnamefont
  {Crouseilles}}, \bibinfo {author} {\bibfnamefont {P.-A.}\ \bibnamefont
  {Hervieux}},\ and\ \bibinfo {author} {\bibfnamefont {G.}~\bibnamefont
  {Manfredi}},\ }\bibfield  {title} {\bibinfo {title} {Quantum hydrodynamic
  model for the nonlinear electron dynamics in thin metal films},\ }\href@noop
  {} {\bibfield  {journal} {\bibinfo  {journal} {Physical Review B}\ }\textbf
  {\bibinfo {volume} {78}},\ \bibinfo {pages} {155412} (\bibinfo {year}
  {2008})}\BibitemShut {NoStop}%
\bibitem [{\citenamefont {Raza}\ \emph {et~al.}(2013)\citenamefont {Raza},
  \citenamefont {Christensen}, \citenamefont {Wubs}, \citenamefont
  {Bozhevolnyi},\ and\ \citenamefont {Mortensen}}]{Raza2013Phys.Rev.B.}%
  \BibitemOpen
  \bibfield  {author} {\bibinfo {author} {\bibfnamefont {S.}~\bibnamefont
  {Raza}}, \bibinfo {author} {\bibfnamefont {T.}~\bibnamefont {Christensen}},
  \bibinfo {author} {\bibfnamefont {M.}~\bibnamefont {Wubs}}, \bibinfo {author}
  {\bibfnamefont {S.~I.}\ \bibnamefont {Bozhevolnyi}},\ and\ \bibinfo {author}
  {\bibfnamefont {N.~A.}\ \bibnamefont {Mortensen}},\ }\bibfield  {title}
  {\bibinfo {title} {Nonlocal response in thin-film waveguides: Loss versus
  nonlocality and breaking of complementarity},\ }\href
  {https://doi.org/10.1103/PhysRevB.88.115401} {\bibfield  {journal} {\bibinfo
  {journal} {Phys. Rev. B}\ }\textbf {\bibinfo {volume} {88}},\ \bibinfo
  {pages} {115401} (\bibinfo {year} {2013})}\BibitemShut {NoStop}%
\bibitem [{\citenamefont {Bozhevolnyi}\ and\ \citenamefont
  {Jung}(2008)}]{Bozhevolnyi2008Opt.Exp.}%
  \BibitemOpen
  \bibfield  {author} {\bibinfo {author} {\bibfnamefont {S.~I.}\ \bibnamefont
  {Bozhevolnyi}}\ and\ \bibinfo {author} {\bibfnamefont {J.}~\bibnamefont
  {Jung}},\ }\bibfield  {title} {\bibinfo {title} {Scaling for gap plasmon
  based waveguides},\ }\href {https://doi.org/10.1364/OE.16.002676} {\bibfield
  {journal} {\bibinfo  {journal} {Opt. Express}\ }\textbf {\bibinfo {volume}
  {16}},\ \bibinfo {pages} {2676} (\bibinfo {year} {2008})}\BibitemShut
  {NoStop}%
\bibitem [{\citenamefont {Smith}\ \emph {et~al.}(2015)\citenamefont {Smith},
  \citenamefont {Stenger}, \citenamefont {Kristensen}, \citenamefont
  {Mortensen},\ and\ \citenamefont {Bozhevolnyi}}]{Smith2015Nanoscale}%
  \BibitemOpen
  \bibfield  {author} {\bibinfo {author} {\bibfnamefont {C.~L.~C.}\
  \bibnamefont {Smith}}, \bibinfo {author} {\bibfnamefont {N.}~\bibnamefont
  {Stenger}}, \bibinfo {author} {\bibfnamefont {A.}~\bibnamefont {Kristensen}},
  \bibinfo {author} {\bibfnamefont {N.~A.}\ \bibnamefont {Mortensen}},\ and\
  \bibinfo {author} {\bibfnamefont {S.~I.}\ \bibnamefont {Bozhevolnyi}},\
  }\bibfield  {title} {\bibinfo {title} {Gap and channeled plasmons in tapered
  grooves: a review},\ }\href {https://doi.org/10.1039/C5NR01282A} {\bibfield
  {journal} {\bibinfo  {journal} {Nanoscale}\ }\textbf {\bibinfo {volume}
  {7}},\ \bibinfo {pages} {9355} (\bibinfo {year} {2015})}\BibitemShut
  {NoStop}%
\bibitem [{\citenamefont {Khalid}\ \emph {et~al.}(2018)\citenamefont {Khalid},
  \citenamefont {{Della Sala}},\ and\ \citenamefont
  {Cirac{\`i}}}]{Khalid2018Opt.Express}%
  \BibitemOpen
  \bibfield  {author} {\bibinfo {author} {\bibfnamefont {M.}~\bibnamefont
  {Khalid}}, \bibinfo {author} {\bibfnamefont {F.}~\bibnamefont {{Della
  Sala}}},\ and\ \bibinfo {author} {\bibfnamefont {C.}~\bibnamefont
  {Cirac{\`i}}},\ }\bibfield  {title} {{\selectlanguage {english}\bibinfo
  {title} {Optical properties of plasmonic core-shell nanomatryoshkas: A
  quantum hydrodynamic analysis}},\ }\href
  {https://doi.org/10.1364/OE.26.017322} {\bibfield  {journal} {\bibinfo
  {journal} {Opt. Express}\ }\textbf {\bibinfo {volume} {26}},\ \bibinfo
  {pages} {17322} (\bibinfo {year} {2018})}\BibitemShut {NoStop}%
\bibitem [{\citenamefont {Khalid}\ and\ \citenamefont
  {Cirac{\`i}}(2019)}]{Khalid2019Photonics}%
  \BibitemOpen
  \bibfield  {author} {\bibinfo {author} {\bibfnamefont {M.}~\bibnamefont
  {Khalid}}\ and\ \bibinfo {author} {\bibfnamefont {C.}~\bibnamefont
  {Cirac{\`i}}},\ }\bibfield  {title} {{\selectlanguage {english}\bibinfo
  {title} {Numerical {{Analysis}} of {{Nonlocal Optical Response}} of
  {{Metallic Nanoshells}}}},\ }\href {https://doi.org/10.3390/photonics6020039}
  {\bibfield  {journal} {\bibinfo  {journal} {Photonics}\ }\textbf {\bibinfo
  {volume} {6}},\ \bibinfo {pages} {39} (\bibinfo {year} {2019})}\BibitemShut
  {NoStop}%
\bibitem [{\citenamefont {Ciracì}\ \emph {et~al.}(2019)\citenamefont
  {Ciracì}, \citenamefont {Jurga}, \citenamefont {Khalid},\ and\ \citenamefont
  {{Della Sala}}}]{Ciraci2019NanoPhotonics}%
  \BibitemOpen
  \bibfield  {author} {\bibinfo {author} {\bibfnamefont {C.}~\bibnamefont
  {Ciracì}}, \bibinfo {author} {\bibfnamefont {R.}~\bibnamefont {Jurga}},
  \bibinfo {author} {\bibfnamefont {M.}~\bibnamefont {Khalid}},\ and\ \bibinfo
  {author} {\bibfnamefont {F.}~\bibnamefont {{Della Sala}}},\ }\bibfield
  {title} {\bibinfo {title} {Plasmonic quantum effects on single-emitter strong
  coupling},\ }\href {https://doi.org/https://doi.org/10.1515/nanoph-2019-0199}
  {\bibfield  {journal} {\bibinfo  {journal} {Nanophotonics}\ }\textbf
  {\bibinfo {volume} {8}},\ \bibinfo {pages} {1821 } (\bibinfo {year} {01 Oct.
  2019})}\BibitemShut {NoStop}%
\bibitem [{\citenamefont {Yang}\ and\ \citenamefont
  {Ding}(2021)}]{Yang2021Photonics}%
  \BibitemOpen
  \bibfield  {author} {\bibinfo {author} {\bibfnamefont {F.}~\bibnamefont
  {Yang}}\ and\ \bibinfo {author} {\bibfnamefont {K.}~\bibnamefont {Ding}},\
  }\bibfield  {title} {\bibinfo {title} {Electron spill-out effect in singular
  metasurfaces},\ }\bibfield  {journal} {\bibinfo  {journal} {Photonics}\
  }\textbf {\bibinfo {volume} {8}},\ \href
  {https://doi.org/10.3390/photonics8050154} {10.3390/photonics8050154}
  (\bibinfo {year} {2021})\BibitemShut {NoStop}%
\bibitem [{\citenamefont {Baghramyan}\ \emph {et~al.}(2021)\citenamefont
  {Baghramyan}, \citenamefont {Della~Sala},\ and\ \citenamefont
  {Cirac\`{\i}}}]{Baghramyan2020}%
  \BibitemOpen
  \bibfield  {author} {\bibinfo {author} {\bibfnamefont {H.~M.}\ \bibnamefont
  {Baghramyan}}, \bibinfo {author} {\bibfnamefont {F.}~\bibnamefont
  {Della~Sala}},\ and\ \bibinfo {author} {\bibfnamefont {C.}~\bibnamefont
  {Cirac\`{\i}}},\ }\bibfield  {title} {\bibinfo {title} {Laplacian-level
  quantum hydrodynamic theory for plasmonics},\ }\href
  {https://doi.org/10.1103/PhysRevX.11.011049} {\bibfield  {journal} {\bibinfo
  {journal} {Phys. Rev. X}\ }\textbf {\bibinfo {volume} {11}},\ \bibinfo
  {pages} {011049} (\bibinfo {year} {2021})}\BibitemShut {NoStop}%
\bibitem [{\citenamefont {Brack}(1993)}]{Brack1993Rev.Mod.Phys.}%
  \BibitemOpen
  \bibfield  {author} {\bibinfo {author} {\bibfnamefont {M.}~\bibnamefont
  {Brack}},\ }\bibfield  {title} {{\selectlanguage {english}\bibinfo {title}
  {The physics of simple metal clusters: Self-consistent jellium model and
  semiclassical approaches}},\ }\href
  {https://doi.org/10.1103/RevModPhys.65.677} {\bibfield  {journal} {\bibinfo
  {journal} {Rev. Mod. Phys.}\ }\textbf {\bibinfo {volume} {65}},\ \bibinfo
  {pages} {677} (\bibinfo {year} {1993})}\BibitemShut {NoStop}%
\bibitem [{\citenamefont {Ajib}\ \emph {et~al.}(2019)\citenamefont {Ajib},
  \citenamefont {Pitelet}, \citenamefont {Poll{\`e}s}, \citenamefont {Centeno},
  \citenamefont {Ajaltouni},\ and\ \citenamefont {Moreau}}]{ajib2019energy}%
  \BibitemOpen
  \bibfield  {author} {\bibinfo {author} {\bibfnamefont {R.}~\bibnamefont
  {Ajib}}, \bibinfo {author} {\bibfnamefont {A.}~\bibnamefont {Pitelet}},
  \bibinfo {author} {\bibfnamefont {R.}~\bibnamefont {Poll{\`e}s}}, \bibinfo
  {author} {\bibfnamefont {E.}~\bibnamefont {Centeno}}, \bibinfo {author}
  {\bibfnamefont {Z.}~\bibnamefont {Ajaltouni}},\ and\ \bibinfo {author}
  {\bibfnamefont {A.}~\bibnamefont {Moreau}},\ }\bibfield  {title} {\bibinfo
  {title} {The energy point of view in plasmonics},\ }\href@noop {} {\bibfield
  {journal} {\bibinfo  {journal} {JOSA B}\ }\textbf {\bibinfo {volume} {36}},\
  \bibinfo {pages} {1150} (\bibinfo {year} {2019})}\BibitemShut {NoStop}%
\bibitem [{\citenamefont {Sakat}\ \emph {et~al.}(2021)\citenamefont {Sakat},
  \citenamefont {Moreau},\ and\ \citenamefont {Hugonin}}]{sakat2021}%
  \BibitemOpen
  \bibfield  {author} {\bibinfo {author} {\bibfnamefont {E.}~\bibnamefont
  {Sakat}}, \bibinfo {author} {\bibfnamefont {A.}~\bibnamefont {Moreau}},\ and\
  \bibinfo {author} {\bibfnamefont {J.-P.}\ \bibnamefont {Hugonin}},\
  }\bibfield  {title} {\bibinfo {title} {Generalized electromagnetic theorems
  for nonlocal plasmonics},\ }\href
  {https://doi.org/10.1103/PhysRevB.103.235422} {\bibfield  {journal} {\bibinfo
   {journal} {Phys. Rev. B}\ }\textbf {\bibinfo {volume} {103}},\ \bibinfo
  {pages} {235422} (\bibinfo {year} {2021})}\BibitemShut {NoStop}%
\bibitem [{\citenamefont {Lema{\^\i}tre}\ \emph {et~al.}(2017)\citenamefont
  {Lema{\^\i}tre}, \citenamefont {Centeno},\ and\ \citenamefont
  {Moreau}}]{lemaitre2017interferometric}%
  \BibitemOpen
  \bibfield  {author} {\bibinfo {author} {\bibfnamefont {C.}~\bibnamefont
  {Lema{\^\i}tre}}, \bibinfo {author} {\bibfnamefont {E.}~\bibnamefont
  {Centeno}},\ and\ \bibinfo {author} {\bibfnamefont {A.}~\bibnamefont
  {Moreau}},\ }\bibfield  {title} {\bibinfo {title} {Interferometric control of
  the absorption in optical patch antennas},\ }\href@noop {} {\bibfield
  {journal} {\bibinfo  {journal} {Scientific Reports}\ }\textbf {\bibinfo
  {volume} {7}},\ \bibinfo {pages} {2941} (\bibinfo {year} {2017})}\BibitemShut
  {NoStop}%
\bibitem [{\citenamefont {Lagarias}\ \emph {et~al.}(1998)\citenamefont
  {Lagarias}, \citenamefont {Reeds}, \citenamefont {Wright},\ and\
  \citenamefont {Wright}}]{Lagarias1998}%
  \BibitemOpen
  \bibfield  {author} {\bibinfo {author} {\bibfnamefont {J.~C.}\ \bibnamefont
  {Lagarias}}, \bibinfo {author} {\bibfnamefont {J.~A.}\ \bibnamefont {Reeds}},
  \bibinfo {author} {\bibfnamefont {M.~H.}\ \bibnamefont {Wright}},\ and\
  \bibinfo {author} {\bibfnamefont {P.~E.}\ \bibnamefont {Wright}},\ }\bibfield
   {title} {\bibinfo {title} {Convergence properties of the nelder-mead simplex
  method in low dimensions.},\ }\href@noop {} {\bibfield  {journal} {\bibinfo
  {journal} {SIAM Journal of Optimization}\ }\textbf {\bibinfo {volume} {9}},\
  \bibinfo {pages} {112} (\bibinfo {year} {1998})}\BibitemShut {NoStop}%
\end{thebibliography}%
\end{document}